\documentclass[%
 reprint,
 amsmath,amssymb,
 aps,
]{revtex4-2}

\usepackage{graphicx}
\usepackage{dcolumn}
\usepackage{bm}
\usepackage{tikz}

\usepackage{mathrsfs}
\usetikzlibrary {arrows.meta}

\definecolor{blue3}{rgb}{0, 0, 1}
\definecolor{blue2}{rgb}{0, 0, 0.7}
\definecolor{blue1}{rgb}{0, 0, 0.4}

\definecolor{red3}{rgb}{1, 0, 0}
\definecolor{red2}{rgb}{0.7, 0, 0}
\definecolor{red1}{rgb}{0.4, 0, 0}

\definecolor{green3}{rgb}{0,1, 0}
\definecolor{green2}{rgb}{0,0.7, 0}
\definecolor{green1}{rgb}{0,0.4, 0}

\definecolor{purple3}{rgb}{1, 0, 1}
\definecolor{purple2}{rgb}{0.7, 0, 0.7}
\definecolor{purple1}{rgb}{0.4, 0, 0.4}

\definecolor{yellow3}{rgb}{0.9, 0.9, 0.45}
\definecolor{yellow2}{rgb}{0.7, 0.7, 0.35}
\definecolor{yellow1}{rgb}{0.4, 0.4, 0.25}

\definecolor{cyan3}{rgb}{0.45, 0.9, 0.9}
\definecolor{cyan2}{rgb}{0.35, 0.7, 0.7}
\definecolor{cyan1}{rgb}{0.25, 0.4, 0.4}

\definecolor{colorFirst}{rgb}{0.368417, 0.506779, 0.709798}
\definecolor{colorSecond}{rgb}{0.880722, 0.611041, 0.142051}
\definecolor{colorThird}{rgb}{0.560181, 0.691569, 0.194885}

\begin{document}

\preprint{APS/123-QED}

\title{A glimpse into the Ultrametric spectrum}

\author{An Huang}%
\affiliation{Department of Mathematics, Brandeis University,
 415 South Street, Waltham, MA 02453, U.S.A.}
\author{Christian Baadsgaard Jepsen}%
\affiliation{School of Physics, Korea Institute for Advanced Study, Seoul 02455, Korea
}
%\date{\today}

\begin{abstract}
The non-relativistic string spectrum is built from integer-spaced energy quanta in such a way that the high-temperature asymptotics, via the Hardy-Ramanujan formula for integer partitions, reduces to standard two-dimensional thermodynamics. Here we explore deformed realizations of this behavior motivated by $p$-adic string theory and Lorentzian versions thereof with a non-trivial spectrum. We study the microstate scaling that results on associating quantum harmonic oscillators to the normal modes of tree-graphs rather than string graphs and observe that Hardy-Ramanujan scaling is not realized. But by computing the eigenvalues of the derivative operator on the $p$-adic circle and by determining the eigenspectrum of the Neumann-to-Dirichlet operator, we uncover a spectrum of exponentially growing energies but with exponentially growing degeneracies balanced in such a way that Hardy-Ramanujan scaling is realized, but modulated with log-periodic fluctuations.

\end{abstract}

\maketitle

\tableofcontents

\section{\label{Sec1}Introduction and Summary}

An important ongoing endeavor in theoretical physics consists in trying to understand the conditions under which string theory provides the unique UV complete quantum theory \cite{Kumar:2009us,Belin:2015hwa,Caron-Huot:2016icg,Sever:2017ylk,Lee:2019wij,Huang:2020nqy,Kaplan:2020ldi,Guerrieri:2021ivu,Arkani-Hamed:2023jwn,Cheung:2024uhn,Cheung:2024obl,Cheung:2025tbr}. The alternatives that have been explored as part of this effort include notions of string theory founded on $p$-adic numbers \cite{volovich1987p,grossman1987p,Freund:1987kt,Zabrodin:1988ep,Huang:2019nog} and $q$-deformations of the usual string \cite{coon1969uniqueness,Figueroa:2022onw,Geiser:2022icl,Chakravarty:2022vrp,Cheung:2022mkw,Geiser:2022exp,Maldacena:2022ckr,Bhardwaj:2022lbz,Cheung:2023adk,Jepsen:2023sia,Li:2023hce,Geiser:2023qqq,Almheiri:2024ayc,Eckner:2024ggx,Rigatos:2024beq,Wang:2024wcc,Belaey:2025kiu}. The study of theories defined over $p$-adic numbers and based on the ultrametric $p$-adic norm derive physical motivation from their applicability to systems based on hierarchical order. Systems organized according to a regular hierarchical pattern exhibit a high degree of symmetry and allow for rigorous derivations, like Freeman Dyson's proof of a second order phase transition \cite{Dyson:1968up}, and they admit their own types of conformal field theories \cite{Melzer:1988he} and holography \cite{Gubser:2016guj,Heydeman:2016ldy}. Within the past five years hierarchical order has been successfully realized in an AMO lab \cite{Periwal:2021eur} as part of an experiment which managed to interpolate between linear and $2$-adic order, after engineering spin couplings inspired by the model put forward in \cite{Gubser:2018bpe,Bentsen:2019rlr}. Moreover, hierarchical behavior can emerge in low energy limits of more general lattice systems \cite{Yan:2023lmj}, \footnote{See also \cite{Capizzi:2022emx} for recent work on quantum hierarchical models and \cite{Banerjee:2026} for recent work on critical phenomena on the Bethe lattice.}. But at the level of the spectrum, $p$-adic string theory in the Freund-Olson formulation is trivial: it contains only the tachyon. This deficiency ultimately hails from the overly simple worldsheet geometry, consisting simply of an infinite regular tree-graph without any separation of spatial and temporal directions. The goal of the present paper is to pave the way for a refinement of the Freund-Olson proposal that too is founded on the hierarchical, ultrametric norm but which captures more features of genuine string theory, including the presence of a tower of states \footnote{See also \cite{Zuniga-Galindo:2022zgm,Zuniga-Galindo:2023ost,Zuniga-Galindo:2023vjj,Zuniga-Galindo:2025otr} for alternative approaches to $p$-adic quantum mechanics and for related notions of statistical field theory \cite{Zuniga-Galindo:2022mhx,Zuniga-Galindo:2023hty}.}.

Unitarity and Lorentz symmetry come together in string theory in a profound and intricate manner that this paper will not seek to generalize. We will content ourselves with the more modest objective of obtaining an ultrametric theory that gives rise to microstates exhibiting the entropy scaling of the non-relativistic string. In the non-relativistic limit, as we review in more detail in Section~\ref{Non-relativistic}, the string spectrum is built from an infinite tower of harmonic oscillators with integer-spaced energy quanta such that the degeneracy of any energy $E=NE_0$ with $N\in\mathbb{N}$ is encapsulated by the integer partition function $\Omega(N)$. By a famous theorem due to Hardy and Ramanujan, this function asymptotes to 
\begin{align}
\label{Hardy-Ramanujan}
\Omega(N)\sim 4^{-1}3^{-1/2}N^{-1}\exp(\pi\sqrt{2N/3})
\end{align}
for large $N$, a behavior that implies an entropy that scales as the square root of energy, in keeping with two-dimensional thermodynamics. The subject of the present investigation, then, is to ascertain whether microstates realizing the same entropy scaling can be realized in the ultrametric setting.

From a discretized perspective, a string can be viewed as an infinite, regular, connected graph of coordination number two. This viewpoint is particularly natural in the context of matrix models, where discretized worldsheets of two-dimensional string theory arises from index contractions of matrix fields \cite{Klebanov:1991qa}. The starting point for our query is to investigate the resultant normal mode spectrum if we generalize the string-graph to a regular tree-graph of coordination number $p+1$ with $p\in \mathbb{N}$, ie. to the infinite graph known as the Bruhat-Tits tree or the Bethe lattice. More precisely, we will look at finite graphs and study their infinite system size limit. In the case of the string graph, this limit is largely insensitive to boundary conditions: irrespectively of whether you impose Dirichlet, Neumann, or period boundary conditions, the infinite system limit produces an integer spectrum. But for finite tree-graphs with $p>1$, the fraction of vertices situated at the boundary constitute a finite fraction of all vertices even in the macroscopic limit, in consequence of which the spectrum changes drastically with the choice of boundary condition. We review the various possibilities in turn:

In Section~\ref{Neumann} we review the spectrum of regular tree-graphs subject to Neumann boundary conditions. In this case a non-trivial spectrum arises in the infinite-system limit, with normal mode frequencies for $p>1$ growing exponentially and also exhibiting an exponential growth in normal mode degeneracy. 

In Section~\ref{Dirichlet} we review the spectrum of regular tree-graphs subject to Dirichlet boundary conditions. The upshot is that for $p>1$, the spectrum becomes infinitely gapped in the macroscopic limit.

In Section~\ref{Periodic} we analyze tree-graphs subject to periodic boundary conditions. While a finite string has but two endpoints to identify with each other, finite tree-graphs possess a multitude of boundary vertices that can be identified with each other in manifold ways. The guiding principle we adopt here is to insist on translational invariance, which in the language of graph theory translates into the condition of vertex transitivity. The graphs we obtain on performing a boundary identification according to this stipulation turn out to be identical to the family of graphs known as Moore graphs. A classification of these graphs exists in the math literature and allows us to rigorously conclude that a macroscopic limit of periodic tree-graphs exists only in the string case $p=1$.

While the first simplistic approach of associating quantum harmonic oscillators to tree-graphs does not achieve the goal of realizing microstates with the desired thermodynamic scaling, the analysis does uncover---particularly in the Neumann case---an important feature that will be part of our subsequent proposed formalism for the ultrametric string, namely the property of exponentially spaced normal modes. This feature too can be found in the above-mentioned attempts to $q$-deform string theory, but unlike the usual focus of those attempts, in the present case the spacings are exponentially increasing rather than decreasing. On its own, the effect of such a more sparse spacing of normal modes is to decrease the entropy. But we encounter too an opposite entropy-increasing effect in the phenomenon of an exponentially growing normal mode degeneracy. Can these two effects be balanced against each other to achieve the desired scaling? In the final part of this paper, we answer the question in the affirmative, subject to an important qualification concerning log-periodic fluctuations. %The spectrum that realizes this balance can be interpreted as arising from a more abstract methodology that seeks to impose an alternate number theoretical metric on the closed string worldsheet.

The closed string can be quantized by performing a foliation of the worldsheet into circles $S^1$. The eigenvalues of the derivative operator acting on the set of functions defined on the circle concur with the integer spectrum of the string. In the case where $p$ is a prime power, there is defined an alternative kind of circle $\mathbb{U}_p$ given by the set of numbers that have unit norm with respect to the ultrametric norm $|\cdot |_p$. The type of derivative that can be defined in this context is the so-called derivative Vladimirov derivative, and in Section~\ref{UltrametricSpectrum} we compute the eigenvalues of this operator when acting the $p$-adic units $\mathbb{U}_p$. Labeling an eigenvalue $\lambda$ and its associated degeneracy $\rho$ as $\lambda_\rho$, so that the spectrum $\mathcal{E}_{(1)}$ of the standard non-relativistic string, up to an overall normalization, can be written as
\begin{align}
\mathcal{E}_{(1)}=\{1_1,\,2_1,\,3_1,\,4_1,\,5_1,\,6_1,\,...\,\}\,,
\end{align}
the eigenvalue sets $\mathcal{E}_{(p)}$ for the first few values of $p$ are given by
\begin{align}
\nonumber
\mathcal{E}_{(2)}&=\{2_1,\,5_2,\,11_4,\,23_8,\,47_{16},\,95_{32},\,...\,\}\,,
\\
\mathcal{E}_{(3)}&=\{1_1,\,5_4,\,17_{12},\,53_{36},\,161_{108},\,485_{324},\,...\,\}\,,
\\ \nonumber
\mathcal{E}_{(5)}&=\{1_3,\,7_{16},\,37_{80},\,187_{400},\,937_{2000},\,4687_{10000},\,...\,\}\,.
\end{align}
For increasing $p$, the spacings between eigenvalues increase, but so too does the growth of degeneracies. In Appendix~\ref{NeumannToDirichlet} we show that up to an overall mass shift, these spectra and their eigenfunctions also furnish the solution to the eigen-problem for the \emph{Neumann-to-Dirichlet} operator on $\mathbb{U}_p$.

In Section~\ref{UltrametricHardRamanujan} we study the thermodynamics that ensue when we associate a quantum harmonic oscillator to each of the eigenmodes in $\mathcal{E}_{(p)}$. The counting of microstates for a given $p$ represents a variation of integer partitioning where only a specific set of exponentially spread integers are permitted in the partition, but where one has to account for the multiplicity of those allowed integers. For example, in the case $p=2$, for the value $N=10$, there are four microstates, $\Omega_{(2)}(10)=4$, associated to the following partitionings:
\begin{align*}
2+2+2+2+2,\hspace{3mm}5_a+5_a,\hspace{3mm}5_a+5_b,\hspace{3mm}5_b+5_b\,.
\end{align*}
Via a saddle point approximation of the thermal partition function, we can determine the large $N$ asymptotics of this counting. In a strict sense, it is impossible to recover the exponentiated square root behavior of \eqref{Hardy-Ramanujan} for the type of spectra we study since the granularity inherent in a sequence of mostly forbidden integers interlaced with sparse, highly degenerate allowed integers induces a modulation in $\Omega_{(p)}(N)$ that is not present in \eqref{Hardy-Ramanujan}. But if we inquire of the averaged asymptotics $\overline{\Omega_{(p)}}$ governing the envelope underlying the modulations, then indeed we recover the desired Hardy-Ramanujan type behavior, e.g., for $p=2$,
\begin{align}
&\hspace{18mm}
\overline{\Omega_{(2)}}(N) \sim \# N^{-a}\,\exp(c\sqrt{N})
\\[2mm] \nonumber
&\text{with}
\hspace{15mm}
a=\frac{5}{4}-\frac{1}{6\log 2}\,,
\hspace{15mm}
c=\frac{\pi}{3}\sqrt{\frac{2}{\log 2}}\,.
\hspace{1mm}
\end{align}
The exponentiated $\sqrt{N}$ behavior holds true for any $p$, and from this behavior we are able to show that the ultrametric spectrum indeed provides a set of microstate energies that at low temperatures produces the standard two-dimensional energy-entropy relation:
\begin{align}
\label{energy-entropy}
\log \Omega_{(p)}\sim \sqrt{E}\,.
\end{align}
We conclude the paper in Section~\ref{Discussion} by discussing the results of this paper and outlining some future directions.

We emphasize that the focus of this paper is on tree-graphs and their continuum limits as we have sought to explore a minimal setting in which to realize the ultrametric behaviour of exponentially spaced and exponentially degenerate spectra.
There are contexts where loop graphs admit of a resummation that reproduce linear Regge behavior, see e.g. \cite{Biswas:2004qu}, and one may envision studying --- at the cost of heightened computational difficulty but with a richer phenomenological pay-off --- graphs that are fractal in some directions and regular square or (hyper-)cubic with loops in other directions. Presumably in such a generalized setting one can realize admixtures of ultrametric and standard, Archimedean behaviors.

\section{\label{Non-relativistic}Non-relativistic String Entropy}
In this section we review the entropy that ensues from the integer-spaced non-relativistic string spectrum. As previously observed in \cite{Tran:2003pf}, evaluating the high temperature partition function of this system via a saddle-point approximation provides a short physics-based derivation of the Hardy-Ramanujan formula for the asymptotics of the integer partition function. We will provide a few extra details in this analysis, including the full sum of non-perturbative corrections to the high-temperature expansion of the string partition function. Subsequently, in Section~\ref{UltrametricHardRamanujan}, we extend this analysis to ultrametric systems with exponentially spaced and exponentially degenerate spectra.

Whether subject to Neumann, Dirichlet, or periodic boundary condition, the classical string exhibits an integer-spaced spectrum of normal modes, and so we take as our starting point the following classical energy values:
\begin{align}
E_k = \varepsilon k\,,
\hspace{10mm}
k\in\mathbb{N}\,.
\end{align}
In the quantum theory we associate a quantum harmonic oscillator to each of these normal modes, resulting in a partition function that is given by
\begin{align}
Z =\,& \prod_{k=1}^\infty \sum_{\ell=0}^\infty e^{-\beta (\ell+\frac{1}{2}) E_k}\,.
\end{align}
Carrying out the sum over $n$ and regulating the zero-point contributions via Riemann zeta regularization, one obtains a partition function given by the reciprocal of the Dedekind eta function,
\begin{align}
Z=\,&e^{\beta \varepsilon/24}\prod_{k=1}^\infty \frac{1}{1-e^{-\beta \varepsilon k}}=
\frac{1}{\eta\big(\frac{i\beta \varepsilon}{2\pi}\big)}\,.
\end{align}
The free energy $F$ is obtained from the partition function via the standard relation
\begin{align}
F =\,& -\frac{1}{\beta}\log Z
\,.
\end{align}
A high temperature expansion of the free energy can be worked out straightforwardly using a known formula relating the logarithm of the Dedekind eta function to the $G_2$ Eisenstein series \footnote{See e.g. equation (10) of \cite{MathworldDedekind}.}:
\begin{align}
\label{derivDedekind}
-4\pi i \frac{d}{d\tau}\log\big[\eta(\tau)\big]
= G_2(\tau)\,.
\end{align}
The Eistenstein series themselves admit of convergent series expressions \footnote{See e.g. equation (3) of \cite{MathworldEisenstein}.}:
\begin{align}
\label{G2k}
G_{2k}(\tau)=2\zeta(2k)+\frac{2(2\pi i)^{2k}}{(2k-1)!}
\sum_{n=1}^\infty \sigma_{2k-1}(n)\,e^{2\pi i n \tau}\,,
\end{align}
where $\zeta(2k)$ is the Riemann-zeta function and $\sigma_{2k-1}(n)$ is the divisor function. Using equations \eqref{derivDedekind} and \eqref{G2k}, one finds the full, convergent high-temperature expansion for the free energy, including non-perturbative corrections,
\begin{align}
\label{Ffull}
F=
-\frac{\pi^2}{6\beta^2 \varepsilon}
\hspace{-0.5mm}-\hspace{-0.5mm}\frac{1}{2\beta}\log(\frac{\beta\varepsilon}{2\pi})
\hspace{-0.5mm}-\hspace{-0.5mm}\frac{1}{\beta}\sum_{n=1}^\infty
\frac{\sigma_1(n)}{n}
\exp\Big(\hspace{-0.5mm}-\hspace{-0.5mm}n \frac{4\pi^2}{\beta\varepsilon}\Big).
\end{align}
To determine the asymptotic density of states, we introduce an entropy function given by
\begin{align}
S(\beta)=\,&\beta E -\beta F\,,
\end{align}
and compute the high temperature density of states $\Omega(E)$ via the entropy, using the standard formula that results from a saddle point evaluation of the path integral:
\begin{align}
\Omega(E) \approx\,& \frac{\exp\big[S(\beta_0)\big]}{\sqrt{2\pi S''(\beta_0)}}\,,
\end{align}
where $\beta_0$ is the solution to the saddle point equation
\begin{align}
0 = S'(\beta_0)\,.
\end{align}
To work out the density of states in a way that will allow us to re-use part of the derivation later, we express the leading terms in the entropy at low temperature as
\begin{align}
S \approx \beta E + \frac{A}{\beta\varepsilon}
+B\log(\beta\varepsilon)-C\,,
\end{align}
where in this specific case the constants $A$, $B$, and $C$ are given by
\begin{align}
\label{ABC}
A = \frac{\pi^2}{6}\,,
\hspace{10mm}
B=\frac{1}{2}\,,
\hspace{10mm}
C= \frac{1}{2}\log(2\pi)\,.
\end{align}
In this notation the saddle point value of the inverse temperature is given by to leading order at small $\beta$ by
\begin{align}
\beta_c = \sqrt{\frac{A}{\varepsilon E}}\,,
\end{align}
and the asymptotic density of states evaluates to
\begin{align}
\label{OmegaE}
\Omega(E)  \approx\,& 
\frac{1}{\varepsilon}\,  \frac{e^{-C}}{2\sqrt{\pi}}
A^{\frac{B}{2}+\frac{1}{4}}
\big(\frac{\varepsilon}{E}\big)^{\frac{B}{2}+\frac{3}{4}}
 \exp\Big[2\sqrt{A\frac{E}{\varepsilon}}\Big]\,,
\end{align}
which for the values given in \eqref{ABC} results in the asymptotic formula
\begin{align}
\Omega(E) \approx\,&
\frac{1}{4\sqrt{3}\,E}
\exp\Big[\pi\sqrt{\frac{2\,E}{3\,\varepsilon}}\Big]\,.
\end{align}
By working out the saddle point expansion to higher orders, one can compute sub-leading corrections, similar to the higher order terms in Hardy's and Ramanujan's asymptotic series formula for the integer partition function. But the method due to Rademacher offers a superior expansion in that it produces a convergent series.

\section{\label{Sec:Tree}Tree Normal Modes}
In this section we review the spectrum of normal modes for the $(p+1)$-regular fractal tree-graph subject to different boundary conditions. Owing to the absence of cycles on the tree-graph, the spectrum is readily solvable, as has been observed at various times in the mathematics, condensed matter, and high energy theory literature, see e.g. \cite{tikhonov2016fractality,aryal2020complete,ostilli2022spectrum,mahan2001energy,solomyak2003spectrum,solomyak2003laplace}.

\begin{figure}[b]
\includegraphics[width=0.8\columnwidth]{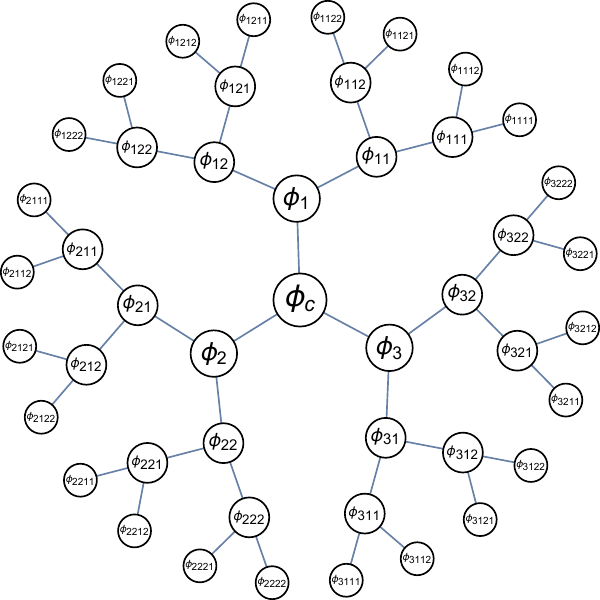}
\caption{\label{fig:tree} The fractal tree for $p=2$ with a cutoff imposed at $L=4$ edges from a central vertex.
}
\end{figure}

The approach we adopt is to solve the equations of motion for a finite, cut-off version of the tree and to subsequently take the infinite system limit. The type of finite tree we consider, sometimes referred to in the literature as the Cayley tree, consists of a central vertex $c$ along with all the vertices situated within a distance of $L$ edges from $c$, as shown on Figure~\ref{fig:tree}. As also shown on the figure, we label the field on the central vertex $\phi_c$, the fields on its neighbors $\phi_i$, and the further neighbor fields $\phi_{ij_1...j_\ell}$ with $\ell\leq L-1$, where $i\in \{1,...,p+1\}$ and $j_1,...,j_\ell \in \{1,...,p\}$. 

The normal mode spectrum is determined by the eigenvalue equation for the graph Laplacian $\square$:
\begin{align}
\label{graphLaplacian}
\lambda \phi_x = \square \phi_x = \sum_{y\sim x}(\phi_x-\phi_y)\,,
\end{align}
where $x \sim y$ indicates that vertices $x$ and $y$ are adjacent. Here we are working in a simplified setup where the graph degrees of freedom $\phi_x\in \mathbb{R}$ are one-dimensional \footnote{A complication present more generally is that a clean separation of oscillations along independent and identical transverse directions does not subsist beyond the special case of chain graphs. In the more general case, the form of the oscillation modes that do not change edge lengths to first order in displacement will depend on the details of the graph's embedding into target space, as studied within the field of rigidity theory.}. We permit ourselves this simplification since the thermodynamic scaling \eqref{energy-entropy} of the non-relativistic string is insensitive to the transverse directions (although the coefficient does depend on the number of transverse directions), whereas we will find that the scaling \eqref{energy-entropy} is not realized for quantum trees.

The task at hand can be viewed as that of determining the normal modes of an oscillating cobweb. It will be important to note that at the classical level, the eigenvalues $\lambda$ of the graph Laplacian are proportional to the \emph{square} of the normal mode frequency $\omega$,
\begin{align}
\label{lambdaOmega}
\lambda \propto \omega^2\,,
\end{align}
with the proportionality constant depending on the precise details of the microscopic model, which will not be important for our purposes of ascertaining thermodynamic scalings.

Under the coordinatization of the tree introduced above, the equations of motion \eqref{graphLaplacian} can be written out more explicitly as
\begin{align}
\nonumber
\lambda \phi_c &= (p+1)\phi_c-\sum_{i=1}^{p+1}\phi_i\,,
\\ 
\lambda \phi_i &=(p+1)\phi_i-\phi_c-\sum_{j=1}^p\phi_{ij}\,,
\label{eoms}
\\ \nonumber
\lambda \phi_{ij_1...j_\ell} &=(p+1)\phi_{ij_1...j_\ell}
-\phi_{ij_1...j_{\ell-1}} - \sum_{j_{\ell+1}=1}^p
\phi_{ij_1...j_{\ell+1}} \,.
\end{align}
For the outermost dynamical fields with $\ell = L-1$, the form of the equation of motion depends crucially on the boundary conditions. And this dependency in turn leads to the normal mode spectra differing drastically depending on the type of boundary conditions. We now proceed to study in turn the spectra ensuing from the different types of boundary conditions, with Subsections~\ref{Neumann} and \ref{Dirichlet} surveying the known spectra resulting from Dirichlet and Neumann boundary conditions respectively, while Subsection~\ref{Periodic} studies the notion of periodic trees. A simple fact that we will make use of in these subsections is that the total number $N_{p,L}$ of nodes on the tree is given by 
\begin{align}
\label{eq:NL}
N_{p,L} = 1 +(p+1)\sum_{n=0}^{L-1}p^n = \frac{p^L(p+1)-2}{p-1}\,.
\end{align}

\subsection{\label{Neumann}Neumann Trees}

Imposing Neumann boundary conditions means that for each edge field $\phi_{ij_1...j_{L-1}}$, we assign the same value to each of its $p$ neighbor fields in the direction away from the centre of the graph: $\phi_{ij_1...j_L}=\phi_{ij_1...j_{L-1}}$. Consequently, the outermost equation of motion is in this case given by
\begin{align}
\lambda \phi_{ij_1...j_{L-1}} &=\phi_{ij_1...j_{L-1}}-\phi_{ij_1...j_{L-2}}\,,
\end{align}

To solve the equations of motion, one can use the equation for the outermost fields $\phi_{ij_1...j_{L-1}}$ to eliminate them in favor of the second-outermost fields $\phi_{ij_1...j_{L-2}}$ provided that $\lambda \neq 1$:
\begin{align}
\phi_{ij_1...j_{L-1}} 
= \frac{1}{1-\lambda}\,
\phi_{ij_1...j_{L-2}}.
\end{align}
Subsequently, one can use the e.o.m. for the second-outermost fields to eliminate them in favor of $\phi_{ij_1...j_{L-3}}$ provided that $p+1-\lambda \neq \frac{p}{1-\lambda}$:
\begin{align}
\phi_{ij_1...j_{L-2}} 
= \frac{1}{h(1-\lambda)}
\,\phi_{ij_1...j_{L-3}}\,,
\end{align}
where we have introduced a function defined by
\begin{align}
\label{hdef}
h(x) \equiv p+1-\lambda-\frac{p}{x}\,.
\end{align}
By iterating this process of eliminating a field in favor of its inner neighbor until one reaches the central field, one obtains a polynomial equation in $\lambda$, whose roots provide values of frequencies for which normal modes exist:
\begin{align}
\label{kappaEq}
\lambda\phi_c =
(p+1)\phi_c-\frac{p+1}{h^{L-1}(1-\lambda)}\phi_c\,,
\end{align}
where $h^{L-1}(1-\lambda)$ signifies the function $h(x)$ applied via function composition $L-1$ times to the value $x=1-\lambda$. The cases where this iterative elimination algorithm cannot be applied to derive equation \eqref{kappaEq} are those where $\lambda$ has a value such that $h^\ell(1-\lambda) = 0$ for any $\ell \in {1,...,L-1}$. Such values of $\lambda$ too yield allowed normal modes, but modes of a different, degenerate kind. When $h^{L-1}(1-\lambda)=0$, the equations of motion demand that $\phi_c=0$, while the equation of motion for $\phi_c$ itself imposes the constraint $\sum_{i=1}^{p+1}\phi_i=0$ so that for each such value of $\lambda$ the normal modes exhibit a $p$-fold degeneracy. Meanwhile, when $h^{L-1-n}(1-\lambda) = 0$, the equations of motion require that all the innermost fields up to and including $\phi_{ij_1...j_{n-1}}$ must vanish, while the non-vanishing fields are subject to the constraints $\sum_{j_n=1}^p\phi_{ij_1...j_n}=0$.

\begin{figure}[t]
\includegraphics[width=\columnwidth]{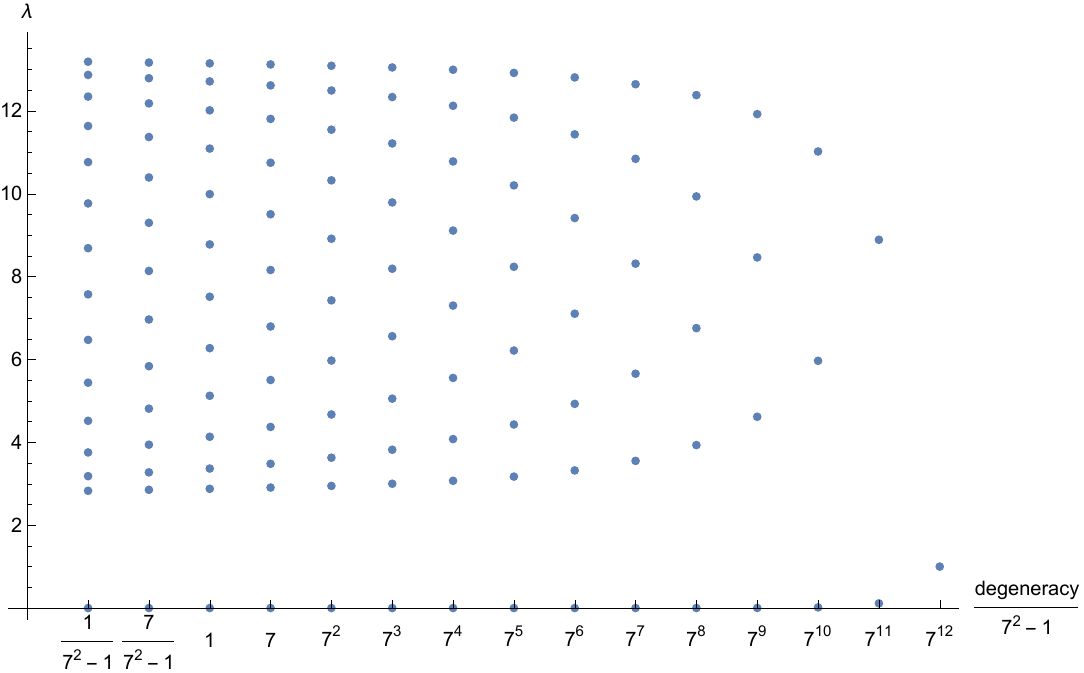}
\caption{\label{fig:14} Plot of the values of $\lambda$ grouped according to their associated degeneracies for the finite fractal tree with $p=7$ and $L=14$. At the lowest degeneracy, the smallest value of $\lambda$ is exactly zero, while at higher degeneracies the smallest values are slightly above zero.
}
\end{figure}

\begin{figure}[t]
\includegraphics[width=0.8\columnwidth]{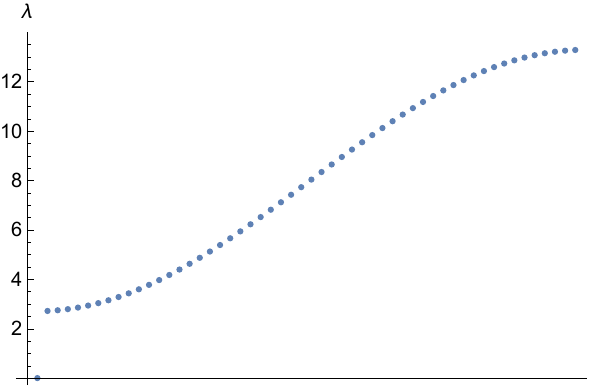}
\caption{\label{fig:53} Plot of the values of $\lambda$ for the 54 solutions to the equation $h^{53}(1-\lambda)=0$ for $p=7$. The value of the first solution is about $1.191\cdot 10^{-45}$.
}
\end{figure}

The upshot is that the finite fractal tree has the following normal modes:
\begin{itemize}
    \item For each of the $L+1$ roots of the polynomial equation
\begin{align}
\label{spectrum1}
p+1-\lambda = \frac{p+1}{h^{L-1}(1-\lambda)}
\end{align}
there is a unique normal mode.
    \item
For each of the $L$ roots of the polynomial equation 
\begin{align}
\label{spectrum2}
h^{L-1}(1-\lambda) = 0
\end{align}
there are $p$ independent normal modes.

    \item
For any $\ell\in \{1,...,L-1\}$ the polynomial equation
\begin{align}
\label{spectrum3}
h^{\ell-1}(1-\lambda) = 0
\end{align}
has $\ell$ roots and for each of these there are $(p^2-1)p^{L-\ell-1}$ independent normal modes.
\end{itemize}
The above set of normal modes and degeneracies provide the complete solution to the equations of motion for Neumann boundary conditions. As a simple check, it is straightforward to verify that we recover the total number of nodes on the tree given in equation~\eqref{eq:NL} when we sum over the multiplicities for each eigenvalue:
\begin{align}
(L+1)+pL+\sum_{\ell=1}^{L-1}\ell(p^2-1)p^{L-\ell-1}=N_{p,L}\,.
\end{align}
Setting $p=1$, the spectrum reduces to the known, non-degenerate normal mode frequencies of the discretized string, as the solutions to \eqref{spectrum1} for $p=1$ are given by
\begin{align}
\label{p1modes1}
\lambda = 4\sin^2\Big(\frac{\pi n}{2L+1}\Big)
\hspace{4mm}
\text{with}
\hspace{4mm}
n\in \{0,1,2,...,L\},
\end{align}
while the $p=1$ solutions to \eqref{spectrum2} are given by
\begin{align}
\label{p1modes2}
\lambda = 4\sin^2\Big(\frac{\pi (n-\frac{1}{2})}{2L+1}\Big)
\hspace{4mm}
\text{with}
\hspace{4mm}
n\in \{1,2,...,L\},
\end{align}
and the normal modes determined by \eqref{spectrum3} fall out of the spectrum as their multiplicities are zero for $p=1$.

Meanwhile, for the trees with $p>1$, the spectra are qualitatively different. To better understand this case, we invite the reader to inspect Figures~\ref{fig:14} and \ref{fig:53}, which plot eigenvalues in the case $p=7$. In Figure~\ref{fig:14}, we plot the allowed eigenvalues as a function of their degeneracies for a system size of $L=14$ and in Figure~\ref{fig:53} we plot the  eigenvalues with $\ell=54$, determined form equation \eqref{p1modes2}. An important feature to observe is that at any fixed degeneracy, as the system size becomes large, all but one of the allowed value of $\lambda$ organize themselves into a smooth sinusoidal curve, with the exception being a single point that approaches zero exponentially in $L$. If one were to take a large system limit $L\rightarrow \infty$ without simultaneously rescaling $\lambda$, one would obtain a system with an infinite ground state degeneracy. But as in the familiar $p=1$ string case, the more suitable large system limit is obtained by performing an appropriate scaling of $\lambda$ with system size as the limit is taken. However, under this rescaling there is a qualitative difference between the cases $p=1$ and $p>1$ in that for $p=1$, the sinusoidal band of $\lambda$ values touches zero and survives in the large system limit, whereas for $p>1$ the band is pushed off to infinity with only the lowest value of $\lambda$ at each degeneracy surviving in the large system limit.

To determine the precise $p>1$ spectrum in the large system limit, it is helpful to rewrite the equation $h^{\ell-1}(1-\lambda)=0$ into the following more explicit form \footnote{While the equation $h^{\ell-1}(1-\lambda)=0$ can be shown to imply \eqref{gequation}, the converse is not true as \eqref{gequation} has two additional roots. However, as the smallest roots is shared by the two equations, the approximation \eqref{kappaApprox} remains valid.}:
\begin{align}
&0=
\big(g(\lambda)+p-1+\lambda\big)\big(p+1-\lambda-g(\lambda)\big)^\ell
+
\nonumber\\[-2.5mm]
\label{gequation}
\\[-2.5mm] \nonumber
&\hspace{6.5mm}
\big(g(\lambda)-p+1-\lambda\big)\big(p+1-\lambda+g(\lambda)\big)^\ell\,,
\\[2mm] \nonumber
& \text{where }\hspace{5mm}g(\lambda) \equiv \sqrt{(p-1)^2-\lambda(2p+2-\lambda)}\,.
\end{align}
Taylor expanding the right-hand side of equation \eqref{gequation} to linear order in $\lambda$ and solving the resultant equation for $\lambda$ yields the approximate solution for the lowest eigenvalue, the only one that survives the large system limit,
\begin{align}
\label{kappaApprox}
\lambda_\ell \approx \frac{(p-1)^2}{p^{\ell+1}+1-(p-1)\ell}\approx (p-1)^2p^{-\ell-1}\,,
\end{align}
with corrections being suppressed exponentially in $\ell$ compared to this expression. This exponential spacing in eigenvalues $\lambda$ leads to an exponential spacing in eigen-frequencies $\omega$ via \eqref{lambdaOmega}. 

To obtain the infinite system limit of the spectrum, we must take the $L\rightarrow \infty$ limit of the eigenvalues $\lambda$ while rescaling them in such a way that the difference between first and second eigenvalues remain constant. For $p>1$, a finite limit value for $\omega^2$ is obtained by taking the $L$ going to infinity limit of the product $(\lambda N_{p,L})$. In this limit, the only eigenvalues of $\lambda$ that do not get pushed off to infinity in the product $(\lambda N_{p,L})$ are the lowest lying eigenvalues approximated by \eqref{kappaApprox}, along with the zero mode, which is present for any value of $L$. Writing $\ell=L-1-n$ with $n\in \mathbb{N}_0$ and taking the $L\rightarrow \infty$ limit, we arrive at the following spectrum of non-zero frequencies $\omega_n$ and associated degeneracies $\rho_n$,
\begin{align}
\label{NeumannSpectrum}
\omega_n&=\omega_0\,p^{n/2}\,,
\\ \nonumber
\rho_0&=p\,,
 &&\rho_n=(p^2-1)p^{n-1}\,,
\hspace{5mm} \text{for }n>0\,,
\end{align}
where the value of the lowest frequency $\omega_0$ depends on the specific microscopic model \footnote{In the simplest model where we consider the fields $\phi_x$ as representing small transverse oscillations of a cobweb stretched along a single spatial direction and think of the vertices as beads of mass $m$, with the edges connecting them serving as springs with a uniform spring constant $k$, we have the relation  $\lambda = m \omega^2/k$. To obtain a finite total mass $M=mN_{p,L}$, the individual bead mass $m$ must scale as $1/N_{p,L}$. Using the fact that the spring constants of springs in parallel are directly additive while those of springs in series are reciprocally additive, one finds that $K = \Big(2\cdot \frac{2}{p+1}\sum_{n=0}^{L-1}p^{-n}\Big)^{-1}k = \frac{p^2-1}{4p(1-p^{-L})}\,k$. For $p>1$, $K$ has a finite limit as $L\rightarrow \infty$ whereas in the case $p=1$ the relation is $K = k/(2L)$. For $p=1$ then, using equations \eqref{p1modes1} and \eqref{p1modes2}, the $L\rightarrow\infty$ limit at fixed $M$ and $K$ then reproduces the familiar string spectrum with integer spacing:
$\omega=\sqrt{\frac{k\lambda}{m}}
=\sqrt{\frac{K2L(2L+1)\lambda}{M}}
\underset{L\rightarrow \infty}{\rightarrow} \pi\sqrt{\frac{K}{M}}n \hspace{3mm}\text{with }n\in\mathbb{N}_0$. For $p>1$ meanwhile, $K$ can be held fixed in the infinite system limit, with the normal mode spectrum determined from the infinite $L$ limit of 
$\omega^2 = \frac{\lambda k}{m} = (\lambda N_{p,L}) \frac{4p(1-p^{-L})}{p^2-1}\frac{K}{M}$, which results in \eqref{NeumannSpectrum} with $\omega_0=2\sqrt{pK/M}$.}.

\subsection{\label{Dirichlet}Dirichlet Trees}

To impose Dirichlet boundary conditions, we set to zero all the non-dynamical outer fields: $\phi_{ij_1...j_L}=0$. This results in the outermost equations of motion taking the form
\begin{align}
\label{DirichletFirst}
\lambda \phi_{ij_1...j_{L-1}} &=(p+1)\phi_{ij_1...j_{L-1}}-\phi_{ij_1...j_{L-2}}\,.
\end{align}
If $\lambda = p+1$, this equation stipulates that the second-outermost fields must vanish: $\phi_{ij_1...j_{L-2}}=0$.
Meanwhile, for $\lambda \neq p+1$, equation \eqref{DirichletFirst} can be straightforwardly solved for the outermost field:
\begin{align}
\lambda \phi_{ij_1...j_{L-1}} &=\frac{1}{p+1-\lambda}\phi_{ij_1...j_{L-2}}\,.
\end{align}
Plugging this solution into the second-outermost equation of motion yields
\begin{align}
\label{DirichletSecond}
\lambda \phi_{ij_1...j_{L-2}} =\,&(p+1)\phi_{ij_1...j_{L-2}}
-\phi_{ij_1...j_{L-3}}
\\ \nonumber
&- \frac{p}{p+1-\lambda}\phi_{ij_1...j_{L-2}}\,.
\end{align}
If $p+1-\lambda=p/(p+1-\lambda)$, this equation stipulates that $\phi_{ij_1...j_{L-3}}=0$. Otherwise \eqref{DirichletSecond} can be used to solve for $\phi_{ij_1...j_{L-2}}$ in terms of $\phi_{ij_1...j_{L-3}}$:
\begin{align}
\phi_{ij_1...j_{L-2}} = \frac{1}{h(p+1-\lambda)}\,\phi_{ij_1...j_{L-3}}\,,
\end{align}
where the function $h(x)$ was defined in \eqref{hdef}. As in the Neumann case, this procedure of solving for fields in terms of their inwardly neighbors---provided the value of $\lambda$ does not require them to vanish---can be iterated all the way to the center of the graph, leading to the equation
\begin{align}
\lambda \phi_c = (p+1)\phi_c-\frac{p+1}{h^{L-1}(p+1-\lambda)}\phi_c\,,
\end{align}
provided that $h^{\ell}(p+1-\lambda)\neq 0$ for $\ell \in \{0,1,...,L-1\}$. In consequence, the spectrum is determined by the values of $\lambda$ that are solutions to the equations 
\begin{align}
\label{hzeroEsq}
    &h^{\ell}(p+1-\lambda)=0\, \text{ for }\ell\in \{0,...,L-1\}\, \text{ or}
\\[1mm]
\label{DirichletFinalEq}
&p+1-\lambda=\frac{p+1}{h^{L-1}(p+1-\lambda)}\,.
\end{align}
The equations \eqref{hzeroEsq} admit a set of closed-form solutions that, in the special case of isotropic normal modes, were also observed in \cite{Ebert:2019src}:
\begin{align}
\label{EbertEq}
    &\lambda_{\ell,n} = p + 1 - 2\,\sqrt{p}\,\cos\big(\frac{n\pi}{\ell+2}\big)\,,
\end{align}
where $n\in\{1,...,\ell+1\}$ is an index that labels different solutions for fixed $\ell$. For our purposes it will also be important to note the multiplicities of each of these solutions:
\begin{align}
\hspace{-2mm}\text{multiplicity}(\lambda_{\ell,n})=\begin{cases}
p\,, \hspace{16mm} \text{for }\ell=L-1
\\[1mm]
(p^2-1)p^{L-\ell-2} \hspace{3mm} \text{otherwise}
\end{cases}
\end{align}
As for the equation \eqref{DirichletFinalEq}, it too admits analytic solutions, each of multiplicity one, given in terms of polynomial roots:
\begin{align}
\label{lambdaLn}
&\lambda_{L,n} = p+1+\sqrt{p}\,\Big(t_n+\frac{1}{t_n}\Big)\,,
\end{align}
where $t_n$ denotes any member of one of the $L+1$ conjugate pairs of roots of the equation
\begin{align}
& 0 = \frac{t_n^{2L+2}(p\,t_n^2-1)+t_n^2-p}{t_n^2-1}\,.
\end{align}
The eigenvalues and multiplicities listed above make up the full set of solutions to the Dirichlet problem on the Cayley tree \footnote{On a technical note, the roots for different values of $\ell$ sometimes overlap so that one must add together the respective multiplicities to get the total degeneracy. E.g., for $p=2$ and $L=6$, the eigenvalue $\lambda=3$ has a total degeneracy of $64$, which comes from a multiplicity-one $\lambda_{L,n}$ eigenvalue, a multiplicity-3 $\lambda_{L-2,n}$ eigenvalue, a multiplicity-12 $\lambda_{L-4,n}$ eigenvalue, and a multiplicity-48 $\lambda_{L-6,n}$ eigenvalue.}. In the limit when $L$ becomes very large, it can be checked that up to an $\mathcal{O}(1/L^3)$ correction, the smallest of the eigenvalues $\lambda_{L,n}$ in \eqref{lambdaLn} tends to $p+1-2\sqrt{p}\cos(\frac{\pi}{L+1})$, which is also the smallest of the eigenvalues $\lambda_{\ell,n}$ in \eqref{EbertEq}. One immediate conclusion then is that in the large system limit, the spectrum becomes trivial for $p>1$. For to obtain the spectrum of eigen-frequencies $\omega$ in the $L\rightarrow \infty$, the eigenvalues $\lambda$ must be multiplied with the system size $N_{p,L}$ to keep energy differences finite. But for Dirichlet boundary conditions, the product $N_{p,L} \lambda $ tends to infinity as $L \rightarrow \infty$, and so the system becomes infinitely gapped \footnote{One could attempt to obtain a non-trivial spectrum through the introduction by hand of a mass term that effectively replaces the graph Laplacian $\square$ in \eqref{graphLaplacian} with $\square -(p+1-2\sqrt{p})$. But this does not lead to a sensible spectrum either. For since $\lambda_{L-\ell,n}-(p+1-\sqrt{p})=\sqrt{p}\frac{n^2\pi^2}{L^2}+\mathcal{O}(1/L^4)$ with no $\ell$ dependence on the right-hand side, this mass shift results in an infinitely degenerate ground state on taking the $L\rightarrow \infty$ limit.}.

\subsection{\label{Periodic}Periodic Trees}

\begin{figure}
{\includegraphics[width=0.4\columnwidth]{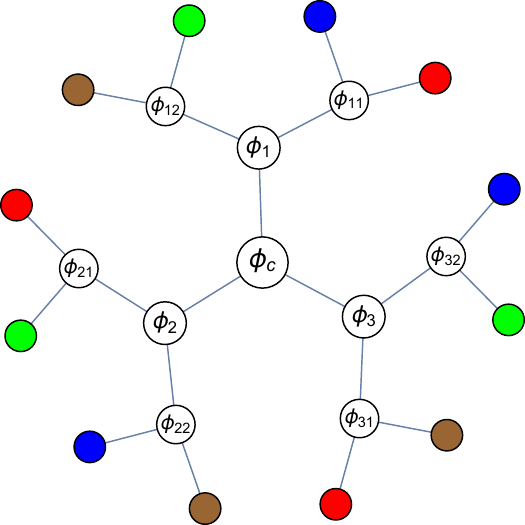}}%
\hspace{10mm}
{\includegraphics[width=0.4\columnwidth]{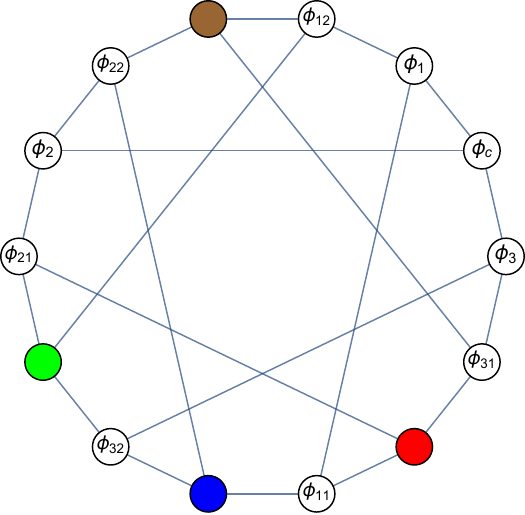}}
\caption{\label{fig:periodicTree}%
By identifying boundary vertices of a finite tree, one can construct a vertex-transitive graph. For this example with $p=2$ and $L=3$, the resulting graph $\mathcal{T}^{(c)}_{2,3}$ is the Heawood graph.
 }%
\end{figure}

Implementing periodic boundary conditions amounts to performing an identification of boundary vertices. But unlike the string case, as the cutoff size $L$ is increased, the boundary points of the tree grow exponentially, leading to a proliferation of possible periodic identifications. However, with a view towards singling out the most symmetric, integrable special case and the closest analog of the closed string, we impose a constraint on the periodic identification by demanding that all vertices of the resulting graph are equivalent. Phrased more precisely, we require that the resultant graph after identifying boundary vertices of the finite tree has the property that any vertex can be mapped to any other via a graph automorphism, ie. the graph is vertex-transitive. Figure~\ref{fig:periodicTree} shows an example of how such a graph is constructed by identifying boundary vertices of a finite tree.

One consequence of the requirement of vertex-transitivity is that the resultant graph must be regular. Hence, the erstwhile boundary vertices must have degree $p+1$, which means that boundary vertices are not identified pairwise but that instead the set of boundary points are partitioned into equivalence classes with $p+1$ members in each class.

A second consequence concerns the cycles in the graph, which are introduced by imposing periodic boundary conditions. In order for a cycle starting at the center vertex to return to the center vertex, the trail must traverse $L$ edges to reach the erstwhile boundary vertices and then must traverse a minimum of $L$ additional edges to return to the center. But since we demand that all vertices are equivalent, we conclude that the minimum length of any cycle, ie. the girth of the graph, is $2L$. Hence, on taking the infinite size limit, cycles of finite length are expunged.

For the values of $p$ and $L$ for which it is possible to construct a vertex-transitive graph by identifying boundary vertices of the finite tree, we will denote the resulting  graph as $\mathcal{T}^{(c)}_{p,L}$. Some members of this family of graphs are known in the literature by special names. In particular, $\mathcal{T}^{(c)}_{2,3}$ is the Heawood graph, $\mathcal{T}^{(c)}_{2,4}$ is the Tutte-Coxeter graph, and $\mathcal{T}^{(c)}_{2,6}$ is the Tutte 12-cage. More generally, a cage in graph theory is defined as a regular graph with the property that there exists no other regular graph with fewer vertices and the same girth.

The total number of vertices $N^{(c)}_{p,L}$ of the graph $\mathcal{T}^{(c)}_{p,L}$ can be determined simply by subtracting from the vertices $N_{p,L}$ of the tree given in Equation~\eqref{eq:NL} the fraction $\frac{p}{p+1}$ of the $(p+1)p^{L-1}$ boundary vertices removed by the identification:
\begin{align}
\label{nodeNumber}
N^{(c)}_{p,L}=
\frac{p^L(p+1)-2}{p-1}-p^L
=\frac{2(p^L-1)}{p-1}\,.
\end{align}
This number $N^{(c)}_{p,L}$ exactly agrees with the known minimum number of vertices for a cage of girth $2L$. It follows that the graph $\mathcal{T}^{(c)}_{p,L}$, whenever such exists, is a cage and what is known in the graph theory literature as a Moore graph \cite{bollobas1998modern}.

It is not hard to see that periodic identification subject to vertex equivalence can produce at most one graph up to isomorphism, so that any answer for fixed $p$ and $L$ is guaranteed to be unique provided it exists. The question is, then, does it exist? As it happens, the precise mathematical question of uniqueness was asked and answered by Robert Singleton in 1966 \cite{singleton1966minimal}. Singleton proved that, for $p$ any integer greater than one, $(p+1)$-regular graphs of girth $2L$ and size $N_{p,L}^{(c)}$ as given in equation \eqref{nodeNumber} do not exist unless $L\in \{1,2,3,4,6\}$.

The case $L=1$ is trivial: $T^{(c)}_{p,1}$ is the $(p+1)-$regular graph with two vertices. 

The case $L=2$ is also quite simple: $T^{(c)}_{p,2}$ is the complete bipartite graph $K_{p+1,p+1}$. 

But for $L=3$, things start to get more complicated. Whenever $T^{(c)}_{p,3}$ exists, it can be identified with the incidence graph of the projective plane, and so $T^{(c)}_{p,3}$ exists if and only if a finite projective plane of order $p$ exists. It follows that $T^{(c)}_{p,3}$ exists whenever $p$ is a prime power, but it is an open problem for what other values of $p$, if any, a projective plane exists; it is known not to exist for $p=6$ and $p=10$ \cite{lam1991search}, but even for $p=12$ the answer is unknown. 

Meanwhile, for $L=4$ and $L=6$, $T^{(c)}_{p,4}$ and $T^{(c)}_{p,6}$ exist for any prime power $p$, as follows from the existence of generalized polygon incidence structures. 

The fact that cages of size $N_{p,L}^{(c)}$ as given in \eqref{nodeNumber} and girth $2L$ do not exist for $L>6$ (except for the $p=1$ case of chain graphs) implies that for fixed $p>1$, there exist only a finite set of periodic trees. Therefore, it is not possible to take an infinite system limit. 

\section{\label{UltrametricSpectrum}The Ultrametric String Spectrum}

In Section~\ref{Sec:Tree} we analyzed the normal mode spectra of infinite regular trees. Under Neumann boundary condition we obtained an exponentially spaced spectrum with an exponentially growing degeneracy, and these features will be essential to the ultrametric Hardy-Ramanujan microstate counting whose derivation constitute a core result of this paper. But the lattice models of Section~\ref{Sec:Tree} and the microstate counting of the quantum trees obtained by associating a quantum harmonic oscillator to each normal mode proves inadequate to achieve our aim of Hardy-Ramanujan scaling. It appears the lattice models are simply too na\"ive. 

The present section adopts a more abstract approach, operating directly in the continuum limit, which will lead us to a spectrum for which Hardy-Ramanujan scaling is realized. This section serves to derive the spectrum, while Section~\ref{UltrametricHardRamanujan} derives the resulting asymptotic microstate counting on associating an infinite set of quantum harmonic oscillators to the spectrum. To obtain the spectrum, we will now restrict $p$ to be a prime number \footnote{The whole analysis can also be straightforwardly extended to the cases when $p$ is a prime power: $p=\mathfrak{p}^n$ with $n\in\mathbb{N}$ and $\mathfrak{p}\in\mathbb{P}$. In this case the set of $p$-adic numbers $\mathbb{Q}_p$ is replaced with the degree $n$ unramified extension of $\mathbb{Q}_p$.}. In this case the boundary of the infinite $(p+1)$-regular tree admits of an identification with the continuous field of $p$-adic numbers $\mathbb{Q}_p$ \footnote{On a technical note, since the boundary of the tree also includes the point at infinite, the boundary space should more precisely be identified with the projective space $\mathbb{P}^1(\mathbb{Q}_p)$.}. In Appendix~\ref{NeumannToDirichlet} we define the Neumann-to-Dirichlet operator for the regular tree-graph and show how, up to an overall shift and a choice of normalization, the eigenspectrum of that operator recovers the spectrum obtained in the present section.

In order to quantize a theory, one must perform a foliation along an isometry. The present section explores a concept of quantization pursuant to a discrete foliation of the tree-boundary according to values of the $p$-adic norm $|\cdot |_p$ so that each slice is given by a set $p^v \mathbb{U}_p$ where $v$ determines the norm of the numbers in a given slice, and $\mathbb{U}_p$ is the set of $p$-adic numbers with unit norm,
\begin{align}
\mathbb{U}_p
=\{\,x\in\mathbb{Q}_p\,\,\big|\,\,|x|_p=1\,\}\,.
\end{align}
States are then determined from an eigen-equation associated to the kinetic term on each slice of the foliation. The eigenfunctions on $\mathbb{U}_p$ can be organized according to their value of a discrete number $\mathscr{C}\in \mathbb{N}_0$ known as the \emph{conductor}, whose precise definition we provide below, after equation~\eqref{lambdapi}. The conductor plays a role analogous to spin. Permutations of branches of the bulk tree can induce isometries of the foliation slices $p^v \mathbb{U}_p$, and for a given boundary function the value of the conductor indicates the sensitivity to permutations of increasingly fine-grained filaments of the tree branches. A large conductor indicates a high sensitivity and an increasing numbers of functions that are mapped to each other under an isometry, in a manner similar to the way in which a large value of spin for a physical particle indicates that the particle sits inside a large irrep under the little group action. And as in quantum field theory solutions to the Dirac equations also solve the Klein-Gordon equation and in string theory the free string equation of motion leads to an infinite tower of higher-spin states, the eigen-functions of the $p$-adic kinetic operator consist of a semi-infinite tower of functions with increasing conductor.

The usual notion of a local derivative operator is absent in the $p$-adic formalism. Instead there exists a bi-local derivative operator known as the Vladimirov derivative. Given a function $\pi:\mathbb{U}_p\rightarrow\mathbb{C}$ on the circle, up to a choice of normalization convention, the Vladimirov derivative $D$ acts as 
\begin{align}
\label{eqDeriv}
D \pi (x) =  \int_{\mathbb{U}_p} dy\,\frac{\pi(x)-\pi(y)}{|y-x|_p^2}\,.
\end{align}
In fact, this is a variation of the original Vladimirov derivative. For more details of this operator and its generalization to the Tate curve, see \cite{HRSW2026}.

A short derivation shows that eigenfunctions are made up of the set of multiplicative characters, ie. the set of functions $\pi:\mathbb{U}_p\rightarrow\mathbb{C}$ that posses the property that for any $z,x\in\mathbb{U}_p$ it holds true that
\begin{align}
\pi(zx) = \pi(z)\,\pi(x)\,.
\end{align}
For when $\pi$ is a multiplicative character, performing a change of variables that sets $y=zx$ and using the fact that $dy=|x|_p dz = dz$, equation \eqref{eqDeriv} can be rewritten as
\begin{align}
D \pi (x) =\,&  \int_{\mathbb{U}_p} dz\,\frac{\pi(x)-\pi(zx)}{|zx-x|_p^2}
\\ =\,& \pi(x) \int_{\mathbb{U}_p} dz\,\frac{1-\pi(z)}{|z-1|_p^2}
\equiv \pi(x)\,\lambda_\pi\,,
\end{align}
where the eigenvalue $\lambda_\pi$ of the Vladimirov derivative associated to the character $\pi$ is given by
\begin{align}
\label{lambdapi}
\lambda_\pi = \int_{\mathbb{U}_p} dz\,\frac{1-\pi(z)}{|z-1|_p^2}\,.
\end{align}
To understand the characters and their eigenvalues it is useful to know that the characters can be classed into families according to their conductors, as mentioned earlier in this section. Each $p$-adic unit $x\in\mathbb{U}_p$ admits of a unique $p$-nary expansion:
\begin{align}
\label{xExpansion}
x = \sum_{n=1}^\infty c_n\,p^{n-1}\,, \hspace{3mm} \text{with}\hspace{3mm} c_n \in \{0,...,p-1\}\,,
\hspace{3mm} c_1 \neq 0\,.
\end{align}
For a character $\pi$ to have conductor $\mathscr{C}$ means that  $\mathscr{C}$ is the smallest non-negative integer such that $\pi(x)$ only depends on the values $c_n$ for $n\leq \mathscr{C}$.
\begin{center}
\textit{$\ast$ Conductor zero $\ast$}
\end{center}
There exists a single unique character of conductor zero, namely the trivial character, for which $\pi(x)=1$ for all $x\in\mathbb{U}_p$. For this character the eigenvalue is immediately seen to equal zero. 
\begin{center}
\textit{$\ast$ Conductor one $\ast$}
\end{center}
For a less trivial example, let us consider the characters of conductor 1, ie. the characters that depend on an argument $x$ only through $c_1\in\{1,...,p-1\}$. We can think of the possible values of $c_1$ as being the elements of the multiplicative group $\mathbb{F}_p^\times$. This group is cyclic with $\varphi(p-1)$ generators, where $\varphi$ is Euler's totient function. Suppose $g$ is one of these generators. Then, up to congruence modulo $p^2$, each possible value of $c_1$ can be expressed as $g^\ell$ with $0\leq \ell \leq p-2$. From this, it follows that we consistently obtain a multiplicative character $\pi$ of conductor 1 by setting, for any $j\in\{1,...,p-2\}$, 
\begin{align}
\label{pi1eq}
\pi(g^\ell) = e^{\frac{2i\pi}{p-1}j\ell}\,.
\end{align}
From the $p - 2$ choices of $j$, the whole set of nontrivial multiplicative characters of conductor 1 is constructed. Let us now compute the eigenvalue \eqref{lambdapi} for such a character. Since the integrand depends only on $z$ through its value of $c_1$, the integral reduces to a finite sum over $p-1$ terms, each carrying a factor of $1/p$ owing to the standard normalization of $p$-adic numbers $\int_{\mathbb{U}_p} dz=\frac{p-1}{p}$,
\begin{align}
\label{conductor1lambda}
\lambda_\pi =\,& \frac{1}{p}
\sum_{\ell=1}^{p-2}\frac{1-e^{\frac{2\pi i}{p-1}j\ell}}{1^2}
= \frac{p-1}{p}\,.
\end{align}
We observe that the characters of conductor 1 are all degenerate since $j$ is absent from the right-hand side of \eqref{conductor1lambda}. In fact it will turn out that all multiplicative characters of any fixed conductor are degenerate.

\begin{center}
\textit{$\ast$ Conductor two $\ast$}
\end{center}
We proceed now to the consider multiplicative characters of conductor two. In this case the eigenvalue integral \eqref{lambdapi} reduces to a sum over $c_1$- and $c_2$-values with each term carrying a measure factor of $1/p^2$. One way to work out the possible values for the characters is to identify the $p(p-1)$ possible values of $\{c_1,c_2\}$ with the multiplicative group of units modulo $p^2$, ie. the group $(\mathbb{Z}/p^2\mathbb{Z})^\times$. This group is a cyclic group of order $p(p-1)$ and, letting $g$ be any generator, the full set of multiplicative characters are given as follows, where $j\in\{1,...,p(p-1)\}$,
\begin{align}
\pi(g) = e^{\frac{2\pi i}{p(p-1)}j}\,.
\end{align}
However, of these $p(p-1)$ characters, the characters with $j\in\{p,2p,...,(p-1)p\}$ actually have conductor one or zero, so we will not consider these but instead focus exclusively on the $p(p-1)-(p-1)=(p-1)^2$ other characters.

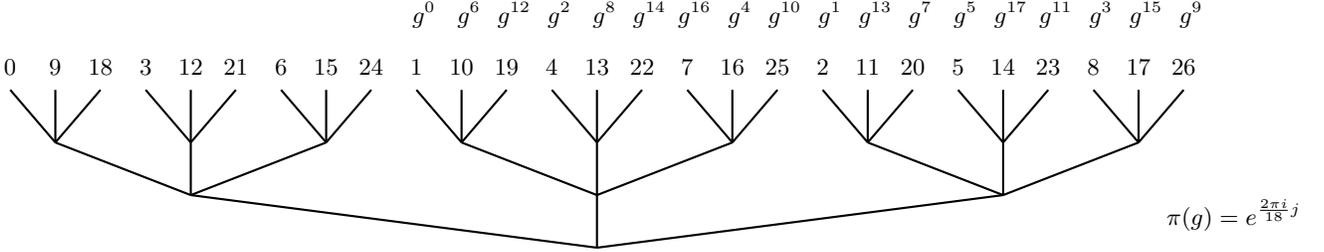
\begin{figure*}
\centering
$
\begin{matrix}
\text{
\scalebox{1}{\begin{tikzpicture}
\node at (0,0.3)  {0};
\node at (0.5,0.3)  {3};
\node at (1,0.3)  {6};
\draw [thick](0.5,-0.7) -- (0,0);
\draw [thick](0.5,-0.7) -- (0.5,0);
\draw [thick](0.5,-0.7) -- (1,0);
\node at (0+1.5,0.3)  {1};
\node at (0.5+1.5,0.3)  {4};
\node at (1+1.5,0.3)  {7};
\node at (0+1.6,1)  {$g^0$};
\node at (0.5+1.6,1)  {$g^2$};
\node at (1+1.6,1)  {$g^4$};
\draw [thick](0.5+1.5,-0.7) -- (0+1.5,0);
\draw [thick](0.5+1.5,-0.7) -- (0.5+1.5,0);
\draw [thick](0.5+1.5,-0.7) -- (1+1.5,0);
\node at (0+3,0.3)  {2};
\node at (0.5+3,0.3)  {5};
\node at (1+3,0.3)  {8};
\node at (3.1,1)  {$g^1$};
\node at (3.6,1)  {$g^5$};
\node at (4.1,1)  {$g^3$};
\draw [thick](0.5+3,-0.7) -- (0+3,0);
\draw [thick](0.5+3,-0.7) -- (0.5+3,0);
\draw [thick](0.5+3,-0.7) -- (1+3,0);
\draw [thick](2,-1.4) -- (0.5,-0.7);
\draw [thick](2,-1.4) -- (2,-0.7);
\draw [thick](2,-1.4) -- (3.5,-0.7);
\end{tikzpicture}}
}
\end{matrix}
\hspace{5mm}
\begin{matrix}
& \pi(g) = e^{\frac{2\pi i}{6}j}
\\[6mm]
&\displaystyle \lambda_\pi=\frac{1}{3^2}\bigg(
\frac{1-\pi(g)^2}{(1/3)^2}+\frac{1-\pi(g)^4}{(1/3)^2}+
\frac{1-\pi(g)^1}{1^2}+\frac{1-\pi(g)^5}{1^2}+\frac{1-\pi(g)^3}{1^2}
\bigg)=\frac{10}{3}
\end{matrix}
$
\caption{
Computation of the eigenvalue for the multiplicative character of conductor two for $p=3$. The 3-adic units $\mathbb{U}_3$ are congruent to 1, 4, 7, 2, 5, and 8 mod 9. These six numbers form a group under multiplication, and we can pick two as a generator $g=2$. The eigenvalue has a four-fold degeneracy as $j$ can assume any of the values 1, 2, 4, and 5.
\label{ConductorTwoTree} 
} 
\end{figure*}

Now, in relating $p$-adic numbers $z\in \mathbb{U}_p$ to $(\mathbb{Z}/p\mathbb{Z})^\times$, it can be seen that when $|z-1|_p=1/p$, the coordinates $c_1$ and $c_2$ have values associated to the elements $g^0$, $g^{p-1}$, $g^{2(p-1)}$,... $g^{(p-1)^2}$. And when $|z-1|_p=1$, the coordinates $\{c_1,c_2\}$ get mapped to the remaining elements of the multiplicative group. From this, one arrives at the following expression for the eigenvalues: 
\begin{align}
\lambda_\pi =\,& \frac{1}{p^2}
\bigg(
\sum_{\ell=1}^{p-1}\frac{1-e^{\frac{2\pi i}{p}j\ell}}{(1/p)^2}
+
\sum_{\ell=1}^{p-2}
\sum_{\ell_2=0}^{p-1}
\frac{1-e^{2\pi i j \frac{\ell+(p-1)\ell_2}{p(p-1)}}}{1^2}
\bigg)
\nonumber
\\
=\,&p+1-\frac{2}{p}
\,.
\end{align}
Again the $j$-dependence drops out, indicating now a $(p-1)^2$-fold degeneracy.  We refer the reader to Figure~\ref{ConductorTwoTree} for an explicit implementation of this computation in the case $p=3$.

\begin{figure*}
\centering
$
\begin{matrix}
\text{
\scalebox{1}{\begin{tikzpicture}
\node at (0,0.3)  {0};
\node at (0.6,0.3)  {9};
\node at (1.2,0.3)  {18};
\draw [thick](0.6,-0.7) -- (0,0);
\draw [thick](0.6,-0.7) -- (0.6,0);
\draw [thick](0.6,-0.7) -- (1.2,0);
\node at (1.8,0.3)  {3};
\node at (2.4,0.3)  {12};
\node at (3,0.3)  {21};
\draw [thick](2.4,-0.7) -- (1.8,0);
\draw [thick](2.4,-0.7) -- (2.4,0);
\draw [thick](2.4,-0.7) -- (3,0);
\node at (3.6,0.3)  {6};
\node at (4.2,0.3)  {15};
\node at (4.8,0.3)  {24};
\draw [thick](4.2,-0.7) -- (3.6,0);
\draw [thick](4.2,-0.7) -- (4.2,0);
\draw [thick](4.2,-0.7) -- (4.8,0);
\draw [thick](2.4,-1.4) -- (0.6,-0.7);
\draw [thick](2.4,-1.4) -- (2.4,-0.7);
\draw [thick](2.4,-1.4) -- (4.2,-0.7);
\node at (0+5.4,0.3)  {1};
\node at (0.6+5.4,0.3)  {10};
\node at (1.2+5.4,0.3)  {19};
\draw [thick](0.6+5.4,-0.7) -- (0+5.4,0);
\draw [thick](0.6+5.4,-0.7) -- (0.6+5.4,0);
\draw [thick](0.6+5.4,-0.7) -- (1.2+5.4,0);
\node at (1.8+5.4,0.3)  {4};
\node at (2.4+5.4,0.3)  {13};
\node at (3+5.4,0.3)  {22};
\draw [thick](2.4+5.4,-0.7) -- (1.8+5.4,0);
\draw [thick](2.4+5.4,-0.7) -- (2.4+5.4,0);
\draw [thick](2.4+5.4,-0.7) -- (3+5.4,0);
\node at (3.6+5.4,0.3)  {7};
\node at (4.2+5.4,0.3)  {16};
\node at (4.8+5.4,0.3)  {25};
\draw [thick](4.2+5.4,-0.7) -- (3.6+5.4,0);
\draw [thick](4.2+5.4,-0.7) -- (4.2+5.4,0);
\draw [thick](4.2+5.4,-0.7) -- (4.8+5.4,0);
\draw [thick](2.4+5.4,-1.4) -- (0.6+5.4,-0.7);
\draw [thick](2.4+5.4,-1.4) -- (2.4+5.4,-0.7);
\draw [thick](2.4+5.4,-1.4) -- (4.2+5.4,-0.7);
\node at (0+10.8,0.3)  {2};
\node at (0.6+10.8,0.3)  {11};
\node at (1.2+10.8,0.3)  {20};
\draw [thick](0.6+10.8,-0.7) -- (0+10.8,0);
\draw [thick](0.6+10.8,-0.7) -- (0.6+10.8,0);
\draw [thick](0.6+10.8,-0.7) -- (1.2+10.8,0);
\node at (1.8+10.8,0.3)  {5};
\node at (2.4+10.8,0.3)  {14};
\node at (3+10.8,0.3)  {23};
\draw [thick](2.4+10.8,-0.7) -- (1.8+10.8,0);
\draw [thick](2.4+10.8,-0.7) -- (2.4+10.8,0);
\draw [thick](2.4+10.8,-0.7) -- (3+10.8,0);
\node at (3.6+10.8,0.3)  {8};
\node at (4.2+10.8,0.3)  {17};
\node at (4.8+10.8,0.3)  {26};
\draw [thick](4.2+10.8,-0.7) -- (3.6+10.8,0);
\draw [thick](4.2+10.8,-0.7) -- (4.2+10.8,0);
\draw [thick](4.2+10.8,-0.7) -- (4.8+10.8,0);
\draw [thick](2.4+10.8,-1.4) -- (0.6+10.8,-0.7);
\draw [thick](2.4+10.8,-1.4) -- (2.4+10.8,-0.7);
\draw [thick](2.4+10.8,-1.4) -- (4.2+10.8,-0.7);
\draw [thick](7.8,-2.1) -- (7.8-5.4,-1.4);
\draw [thick](7.8,-2.1) -- (7.8,-1.4);
\draw [thick](7.8,-2.1) -- (7.8+5.4,-1.4);
\node at (5.5,1)  {$g^0$};
\node at (6.1,1)  {$g^6$};
\node at (6.7,1)  {$g^{12}$};
\node at (7.3,1)  {$g^2$};
\node at (7.9,1)  {$g^8$};
\node at (8.5,1)  {$g^{14}$};
\node at (9.1,1)  {$g^{16}$};
\node at (9.7,1)  {$g^4$};
\node at (10.3,1)  {$g^{10}$};
\node at (10.9,1)  {$g^1$};
\node at (11.5,1)  {$g^{13}$};
\node at (12.1,1)  {$g^7$};
\node at (12.7,1)  {$g^5$};
\node at (13.3,1)  {$g^{17}$};
\node at (13.9,1)  {$g^{11}$};
\node at (14.5,1)  {$g^{3}$};
\node at (15.1,1)  {$g^{15}$};
\node at (15.7,1)  {$g^9$};
\end{tikzpicture}}
}
\begin{matrix}
\hspace{-7mm}
\begin{matrix}
\\[10mm]
\pi(g)=e^{\frac{2\pi i}{18}j}    
\end{matrix}
\\[28mm]
\end{matrix}
\\[-12mm]
\begin{matrix}
\displaystyle \lambda_{\pi}=\frac{1}{3^3}\bigg(
\frac{1-\pi(g)^6}{(1/9)^2}+
\frac{1-\pi(g)^{12}}{(1/9)^2}+
\frac{1-\pi(g)^2}{(1/3)^2}+
\frac{1-\pi(g)^8}{(1/3)^2}+
\frac{1-\pi(g)^{14}}{(1/3)^2}+
\frac{1-\pi(g)^{16}}{(1/3)^2}+
\frac{1-\pi(g)^4}{(1/3)^2}+
\frac{1-\pi(g)^{10}}{(1/3)^2}+
\textcolor{white}{OOOOO}
\\[3mm]
\displaystyle
\frac{1-\pi(g)^1}{1^2}+
\frac{1-\pi(g)^{13}}{1^2}+
\frac{1-\pi(g)^7}{1^2}+
\frac{1-\pi(g)^5}{1^2}+
\frac{1-\pi(g)^{17}}{1^2}+
\frac{1-\pi(g)^{11}}{1^2}+
\frac{1-\pi(g)^3}{1^2}+
\frac{1-\pi(g)^{15}}{1^2}+
\frac{1-\pi(g)^9}{1^2}
\bigg)=\frac{34}{3}
\end{matrix}
\end{matrix}
$
\caption{
Computation of the eigenvalue for the multiplicative character of conductor three for $p=3$. Modulo 27, the 3-adic units $\mathbb{U}_3$ are congruent to the set of numbers $\{1,10,19,4,13,22,7,16,25,2,11,20,5,14,23,8,17,26\}$ and these numbers form a cyclic group under multiplication mod 27. There is an 12-fold degeneracy as $j$ can assume any value in the set $\{1,2,4,5,7,8,10,11,13,14,16,17\}$. 
\label{ConductorThreeTree} 
} 
\end{figure*}

\begin{center}
\textit{$\ast$ Conductor three $\ast$}
\end{center}

Our list of examples is beginning to grow lengthy, but we will include also the case of conductor 3 before proceeding to the general case, as this last example will make it easier to parse the general computation. As the computation works out a little differently depending on whether $p$ is two or an odd prime, we will consider the two cases separately. 

Let us first take $p$ to be an odd prime. The set of possible values for $\{c_1,c_2,c_3\}$ can be identified with the multiplicative group $(\mathbb{Z}/p^3\mathbb{Z})^\times$, which is again cyclic and which has order $p^2(p-1)$. The values of the multiplicative characters when evaluated on any generator $g$ are given by
\begin{align}
\pi(g) = e^{\frac{2\pi i}{p^2(p-1)}j}\,,
\end{align}
with $1\leq j \leq p^2(p-1)$, where the characters with $j\in\{p,...,(p-1)p^2\}$ have conductor zero, one, or two, and so will not be consider here. Mapping the elements of $z\in\mathbb{U}_p$ to $(\mathbb{Z}/p^3\mathbb{Z})^\times$ via their coordinates $\{c_1,c_2,c_3\}$, one finds that 
\begin{itemize}
    \item $|z-1|_p=p^{-2}$ when $z$ is mapped to  $g^{p(p-1)\ell}$ with $\ell\in\{0,...,p-1\}$.
    \item $|z-1|_p=p^{-1}$ when $z$ is mapped to $g^{(p-1)(\ell_1+p\ell_2)}$ with $\ell_1\in\{1,...,p-1\}$ and $\ell_2\in\{0,...,p-1\}$.
    \item $|z-1|_p=1$ when $z$ is mapped to any other element of $(\mathbb{Z}/p^3\mathbb{Z})^\times$.
\end{itemize}
From this, the $(p-1)^2p$ eigenvalues for the characters of conductor three evaluate to the following:
\begin{align}
\lambda_\pi =\,& \frac{1}{p^3}
\bigg(
\sum_{\ell=1}^{p-1}\frac{1-e^{\frac{2\pi i}{p}j\ell}}{(1/p^2)^2}
+
\sum_{\ell_1=1}^{p-1}
\sum_{\ell_2=0}^{p-1}
\frac{1-e^{2\pi i j \frac{\ell_1+p\ell_2}{p^2}}}{(1/p)^2}
\nonumber
\\
&
\hspace{7mm}
+
\sum_{\ell=1}^{p-2}
\sum_{\ell_2=0}^{p-1}
\sum_{\ell_3=0}^{p-1}
\frac{1-e^{2\pi i j \frac{\ell+(p-1)\ell_2+p(p-1)\ell_3}{p(p-1)}}}{1^2}
\bigg)
\nonumber
\\
=\,&p^2+p-\frac{2}{p}
\,.
\end{align}
See Figure~\ref{ConductorThreeTree} for an explicit implementation of the calculation for $p=3$.

Let us now consider the case when $p=2$. The group $(\mathbb{Z}/2^3\mathbb{Z})^\times$ is isomorphic to $\mathbb{Z}_2\times \mathbb{Z}_2$. Labeling the generators of the two copies of $\mathbb{Z}_2$ as $g_1$ and $g_2$, the character of a group element $(g_1)^{n_1}\times(g_2)^{n_2}$ with $n_1,n_2\in\{0,1\}$ is given by $\pi(g_1)^{n_1}\pi(g_2)^{n_2}$ where $\pi(g_1),\pi(g_2)\in\{-1,1\}$. The character with $\pi(g_1)=\pi(g_2)=1$ has conductor zero and the character with $\pi(g_1)=\pi(g_2)=-1$ has conductor two. The remaining two characters with $\pi(g_1)=-\pi(g_2)=\pm 1$ have conductor three, and their eigenvalues are identical, being given by
\begin{align}
\lambda_\pi=\frac{1}{2^3}
\Big(\frac{1-(-1)}{(1/2^2)^2}+\frac{1-1}{(1/2)^2}+\frac{1-(-1)}{(1/2)^2}\Big)=5\,.
\end{align}

\begin{figure*}
\centering
$
\begin{matrix}
\text{
\scalebox{1}{\begin{tikzpicture}
\node at (0,0.3)  {0};
\node at (1,0.3)  {8};
\node at (2,0.3)  {4};
\node at (3,0.3)  {12};
\node at (4,0.3)  {2};
\node at (5,0.3)  {10};
\node at (6,0.3)  {6};
\node at (7,0.3)  {14};
\node at (8,0.3)  {1};
\node at (8,1)  {$g_1^0g_2^0$};
\node at (9,0.3)  {9};
\node at (9,1)  {$g_1^0g_2^2$};
\node at (10,0.3)  {5};
\node at (10,1)  {$g_1^1g_2^1$};
\node at (11,0.3)  {13};
\node at (11,1)  {$g_1^1g_2^3$};
\node at (12,0.3)  {3};
\node at (12,1)  {$g_1^0g_2^1$};
\node at (13,0.3)  {11};
\node at (13,1)  {$g_1^0g_2^3$};
\node at (14,0.3)  {7};
\node at (14,1)  {$g_1^1g_2^0$};
\node at (15,0.3)  {15};
\node at (15,1)  {$g_1^1g_2^2$};
\draw [thick](0.5,-0.5) -- (0,0);
\draw [thick](0.5,-0.5) -- (1,0);
\draw [thick](2.5,-0.5) -- (2,0);
\draw [thick](2.5,-0.5) -- (3,0);
\draw [thick](4.5,-0.5) -- (4,0);
\draw [thick](4.5,-0.5) -- (5,0);
\draw [thick](6.5,-0.5) -- (6,0);
\draw [thick](6.5,-0.5) -- (7,0);
\draw [thick](8.5,-0.5) -- (8,0);
\draw [thick](8.5,-0.5) -- (9,0);
\draw [thick](10.5,-0.5) -- (10,0);
\draw [thick](10.5,-0.5) -- (11,0);
\draw [thick](12.5,-0.5) -- (12,0);
\draw [thick](12.5,-0.5) -- (13,0);
\draw [thick](14.5,-0.5) -- (14,0);
\draw [thick](14.5,-0.5) -- (15,0);
\draw [thick](1.5,-1) -- (0.5,-0.5);
\draw [thick](1.5,-1) -- (2.5,-0.5);
\draw [thick](5.5,-1) -- (4.5,-0.5);
\draw [thick](5.5,-1) -- (6.5,-0.5);
\draw [thick](9.5,-1) -- (8.5,-0.5);
\draw [thick](9.5,-1) -- (10.5,-0.5);
\draw [thick](13.5,-1) -- (12.5,-0.5);
\draw [thick](13.5,-1) -- (14.5,-0.5);
\draw [thick](3.5,-1.5) -- (1.5,-1);
\draw [thick](3.5,-1.5) -- (5.5,-1);
\draw [thick](11.5,-1.5) -- (9.5,-1);
\draw [thick](11.5,-1.5) -- (13.5,-1);
\draw [thick](7.5,-2) -- (3.5,-1.5);
\draw [thick](7.5,-2) -- (11.5,-1.5);
\end{tikzpicture}}
}
\begin{matrix}
\hspace{-7mm}
\begin{matrix}
\\[5mm]
\pi(g_1)=e^{i\pi j_1}
\textcolor{white}{O}
\\[2mm]
\pi(g_2)=e^{i\pi j_2/2}
\end{matrix}
\\[28mm]
\end{matrix}
\\[-12mm]
\begin{matrix}
\displaystyle \lambda_\pi=
\frac{1}{2^4}
\bigg(
\frac{1-\pi(g_2)^2}{(1/8)^2}+
\frac{1-\pi(g_1)\pi(g_2)}{(1/4)^2}+
\frac{1-\pi(g_1)\pi(g_2)^3}{(1/4)^2}+
\frac{1-\pi(g_2)}{(1/2)^2}+
\frac{1-\pi(g_2)^3}{(1/2)^2}+
\frac{1-\pi(g_1)}{(1/2)^2}+
\frac{1-\pi(g_1)\pi(g_2)^2}{(1/2)^2}
\bigg)
=11
\end{matrix}
\end{matrix}
$
\caption{
Computation of the eigenvalue for the multiplicative characters of conductor four for $p=2$. Modulo 16, the 2-adic units $\mathbb{U}_2$ are congruent to the set of numbers $\{1,9,5,13,3,11,7,15\}$ and these numbers form a group isomorphic to $\mathbb{Z}_2\times\mathbb{Z}_4$ under multiplication mod 16. We can choose 7 as a generator of the $\mathbb{Z}_2$ factor and 3 as a generator of the $\mathbb{Z}_4$ factor. There is a four-fold degeneracy as $\lambda_\pi$ evaluates to the same value for any $j_1\in\{1,2\}$ and $j_2\in\{1,3\}$. 
\label{ConductorFour} 
} 
\end{figure*} 

\begin{center}
\textit{$\ast$ Conductor $n$ $\ast$ }
\end{center}

After the warm-up of working out the eigenvalues for characters of conductors one through three, it is not difficult to establish the general formula for conductor $n$ by the same method. Again we will treat the even and odd primes separately.

Consider first the case when $p$ is an odd prime. By mapping $p$-adic numbers $z\in\mathbb{U}_p$ to the cyclic group $(\mathbb{Z}/p^n\mathbb{Z})^\times$, one finds that the multiplicative characters of conductor $n$ can be labelled by an integer $j$ running from 1 to $p^{n-1}(p-1)$ and not divisible with $p$, and the character eigenvalues can be determined from the formula
\begin{align}
&\hspace{18mm}\lambda_\pi=\frac{1}{p^n}
\bigg(
\sum_{\ell=1}^{p-1}
\frac{1-\exp(\frac{2\pi ij}{p}\ell)}{(1/p^{n-1})^2}
+
\\ \nonumber
&
\sum_{k=2}^{n-1}
\sum_{\ell_1=1}^{p-1}
\sum_{\ell_2,...,\ell_k=0}^{p-1}
\frac{1-\exp(\frac{2\pi ij}{p^k}\sum_{h=1}^k p^{h-1}\ell_h)}{(1/p^{n-k})^2}
+
\\ \nonumber
&
\sum_{\ell=1}^{p-2}
\sum_{\ell_2,...,\ell_n=0}^{p-1}
\hspace{-3.3mm}
\frac{1\hspace{-0.5mm}-\hspace{-0.3mm}\exp\hspace{-0.5mm}\big[\frac{2\pi ij}{(p-1)p^{n-1}}(\ell\hspace{-0.5mm}+\hspace{-0.5mm}(p\hspace{-0.5mm}-\hspace{-0.5mm}1)\sum_{h=2}^n p^{h-2}\ell_h)\big]}{1^2}
\bigg)
\\
&\hspace{20mm}\textcolor{white}{\lambda}=(p+1)p^{n-2}-\frac{2}{p}\,.
\end{align}

Now we turn to the case $p=2$. The group $(\mathbb{Z}/2^n \mathbb{Z})^\times$ is isomorphic to $\mathbb{Z}_2\times \mathbb{Z}_{2^{n-2}}$. Let $g_1$ be the generator of the $\mathbb{Z}_2$ factor. There are $2^{n-2}$ characters of conductor $n$, which can be enumerated by assigning to $\pi(g_1)$ a value of plus or minus one and by picking any generator $g_2$ out of the $2^{n-3}$ generators of $\mathbb{Z}_{2^{n-2}}$ and setting $\pi(g_2)=e^{\frac{\pi i}{2^{n-3}}}$. These characters all have the same eigenvalue given by
\begin{align}
\lambda_\pi
=\,&\frac{1}{2^n}
\bigg(
\sum_{\ell=1}^{n-2}\sum_{k=1}^{2^{\ell-1}}
\frac{1-\exp(\frac{2\pi i(2k-1)}{2^\ell})}{(1/2^{n-l})^2}
\\
\nonumber
&\hspace{9mm}+\sum_{\ell=1}^{2^{n-2}}
\frac{1-\exp(\frac{\pi i \ell}{2^{n-3}})}{(1/2)^2}
\bigg)=3\cdot 2^{n-2}-1\,.
\end{align}
See Figure~\ref{ConductorFour} for an explicit implementation of the calculation for the characters of conductor four. We observe that though the derivation was not entirely the same, the eigenvalues and degeneracies for $p=2$ match the values obtained by setting $p$ to two in the final $p$-dependent answers for the eigenvalues and degeneracies in the odd prime case.

We have now determined the full set of eigenvalues and degeneracies for the Vladirov derivative acting on functions on the $p$-adic units $\mathbb{U}_p$. Up to an overall dimensionful constant that we will label $\varepsilon$, it is this set of eigenvalues that we denominate as the ultrametric string spectrum. The conclusion of the present section is that this spectrum is given by the following set of energy values $E_n$ and associated degeneracies $\rho_n$:
\begin{align}
\nonumber
E_n &= \varepsilon\Big(\frac{p+1}{p^2}p^n-\frac{2}{p}\Big)\,,
\hspace{5mm} n\in \mathbb{N}\,.
\\[2mm] 
\rho(1) &= p-2\,,
\label{ultSpectrum}
\\[1mm]
\rho(n) &=(p-1)^2p^{n-2} \text{ for }
 n>1\,. \nonumber
\end{align}

\section{\label{UltrametricHardRamanujan}Ultrametric String Entropy}

In Section~\ref{UltrametricSpectrum} we derived what we dubbed the ultrametric string spectrum. The spectrum displays a qualitative similarity to the normal modes of the infinite fractal tree with Neumann boundary conditions reviewed in \ref{Neumann}  insofar as these spectra consist of exponentially spaced energy levels with exponentially growing degeneracies. While these spectra are purely classical, the present section studies the quantum mechanical system that arises on promoting each energy level to the energy quantum of a quantum harmonic oscillator. 

In direct analogy with the non-relativistic string reviewed in Section~\ref{Non-relativistic}, the infinite set of quantum oscillators gives rise to a multiplicity of possible microstates at any given energy. At a low value of energy, the exact total multiplicity can be determined via the combinatorial problem of counting all possible ways of adding together oscillator quanta to obtain the given energy. The objective of the present section is to compute the asymptotic behavior of the multiplicity at large energies and thereby determine the relation between energy and entropy at large temperature.

Before we write down the partition function for the quantum system, let us address the question of zero point energy. In the usual string case, as we saw in Section~\ref{Non-relativistic}, the divergent zero-point contribution admits of Riemann zeta regularization, and including this contribution results in the partition function being precisely a reciprocal Dedekind eta function. Moreover, in the case of 24 transverse directions, the regulated energy offset produced by the zero-point energies shifts the ground state into the tachyonic value of the bosonic string:
\begin{align*}
24\cdot \sum_{n=1}^\infty \frac{n}{2} = 12\,\zeta(-1) = -1\,.
\end{align*}
In the case of the ultrametric spectrum, the zero-point contribution to the partition function is again divergent. The divergence manifests itself in the form of a geometric series with a common ratio greater than one. Arguably, the most natural regulator consist in replacing such a divergent series with the resummed answer, which amounts to performing an analytic continuation in the value of $p$, starting from complex values with $|p|<1$ and then continuing to the desired prime value. Under this regularization scheme, the total contribution from the 1/2 terms of the quantum harmonic oscillators for the precise energies and degeneracies given in \eqref{ultSpectrum} turn out to vanish exactly,
\begin{align}
\label{zeroPointEnergy}
& \hspace{28mm}
\sum_{n\in\mathbb{N}}
\rho(n) E_n
=
\\ \nonumber
&(p-2)\varepsilon\frac{p-1}{p}
+\sum_{n=2}^\infty(p-1)^2
p^{n-2}
\varepsilon
\Big(\frac{p+1}{p^2}p^n-\frac{2}{p}\Big)
=0\,.
\end{align}
In the absence of a zero-point contribution, the partition function at inverse temperature $\beta$ is given by
\begin{align}
Z = 
 \hspace{-0.5mm}
\prod_{n=1}^\infty 
 \hspace{-0.5mm}
 \bigg(\sum_{\ell=1}^\infty
e^{-\beta \ell E_n}\bigg)^{\rho(n)}
 \hspace{-2mm}
 = 
 \hspace{-0.5mm}
 \prod_{n=1}^\infty 
  \hspace{-0.5mm}
 \bigg(\frac{1}{1-e^{-\beta E_n}}\bigg)^{\rho(n)}
  \hspace{-3mm}.
  \end{align}
The free energy $F$ and entropy $S$ follow from the partition function and energy $E$ according to the standard relations
\begin{align}
F =-\frac{1}{\beta}\log Z\,,
\hspace{10mm}
S=\beta E +\log Z\,.
\end{align}
Our present goal is to understand the large temperature asymptotics of these functions. For the usual string, we saw in equation \eqref{Ffull} that it was possible to determine all terms in the small $\beta$ expansion of the free energy. The same turns out to be true in this ultrametric setting, as we show in Appendix~\ref{Fderiv}. 

\begin{figure*}
\centering
$
\begin{matrix}
\text{
\includegraphics[width=0.6\columnwidth]{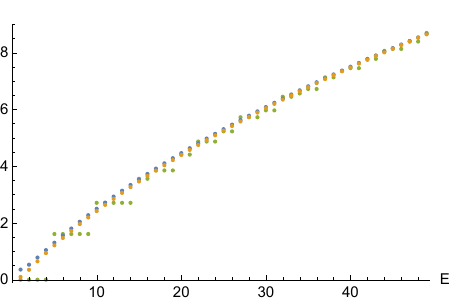}}
\end{matrix}
\hspace{7mm}
\begin{matrix}
\text{
\includegraphics[width=0.6\columnwidth]{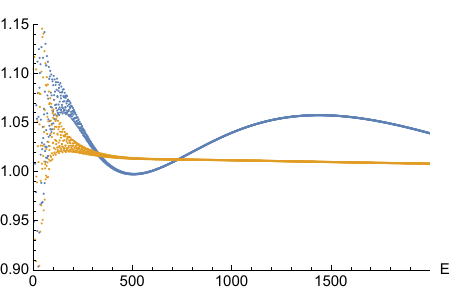}}
\end{matrix}
\hspace{7mm}
\begin{matrix}
\text{
\includegraphics[width=0.6\columnwidth]{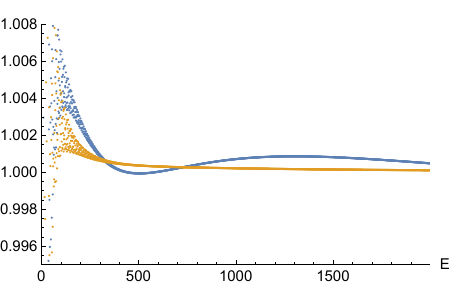}}
\end{matrix}
$
\\
$
\begin{matrix}
\text{
\scalebox{1}{\begin{tikzpicture}
\filldraw[colorThird] (0,0) circle (1.3pt);
\node at (0.8,0)  {$\log\Omega_\text{exact}$};
\filldraw[colorFirst] (1.8,0) circle (1.3pt);
\node at (2.68,0)  {$\log\overline{\Omega_\text{asymp}}$};
\filldraw[colorSecond] (3.7,0) circle (1.3pt);
\node at (4.58,0)  {$\log\Omega_\text{asymp}$};
\end{tikzpicture}}}
\end{matrix}
\hspace{17mm}
\begin{matrix}
\text{
\scalebox{1}{\begin{tikzpicture}
\filldraw[colorFirst] (0,0) circle (1.3pt);
\node at (0.7,0)  {$\displaystyle\frac{\overline{\Omega_\text{asymp}}}{\Omega_\text{exact}}$};
\filldraw[colorSecond] (1.8,0) circle (1.3pt);
\node at (2.5,0)  {$\displaystyle\frac{\Omega_\text{asymp}}{\Omega_\text{exact}}$};
\end{tikzpicture}}}
\end{matrix}
\hspace{24mm}
\begin{matrix}
\text{
\scalebox{1}{\begin{tikzpicture}
\filldraw[colorFirst] (0,0) circle (1.3pt);
\node at (1,0)  {$\displaystyle\frac{\log\overline{\Omega_\text{asymp}}}{\log\Omega_\text{exact}}$};
\filldraw[colorSecond] (2.1,0) circle (1.3pt);
\node at (3.1,0)  {$\displaystyle\frac{\log\Omega_\text{asymp}}{\log\Omega_\text{exact}}$};
\end{tikzpicture}}}
\end{matrix}
$
\\[-3mm]
$
\begin{matrix}
\text{
\includegraphics[width=0.6\columnwidth]{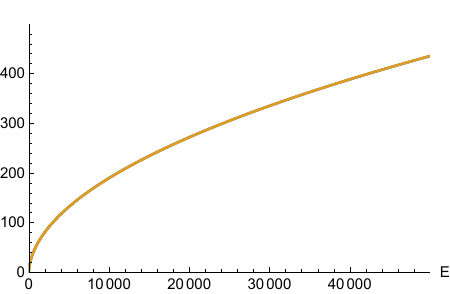}}
\end{matrix}
\hspace{7mm}
\begin{matrix}
\text{
\includegraphics[width=0.6\columnwidth]{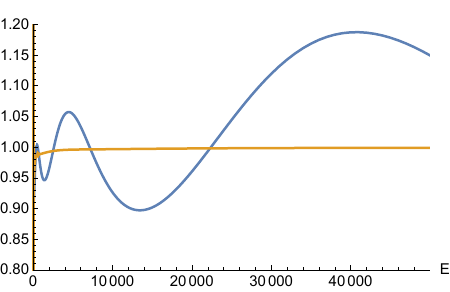}}
\end{matrix}
\hspace{7mm}
\begin{matrix}
\text{
\includegraphics[width=0.6\columnwidth]{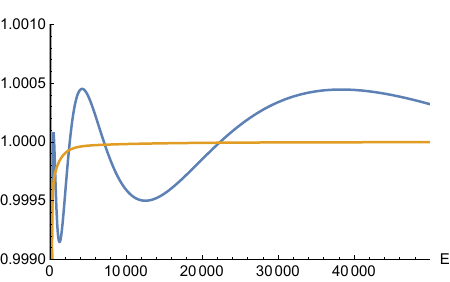}}
\end{matrix}
$
\caption{Plots of the exact ultrametric degeneracies along with the averaged and unaveraged asymptotic values for $p=3$ and $\varepsilon=3/2$. The two plots in each of the three columns differ only by their plot regions, with the lower half displaying a wider range of energies. The exact degeneracy oscillates around a mean asymptotic value, as correctly captured by the unaveraged asymptotic function. The three sets of datapoints on the lower left-hand figure almost coincide, giving the impression of a single curve. \label{HRplots} 
} 
\end{figure*}

At high temperature, $F$ scales as $\beta^{-2}$ to leading order, just as in the usual string case, but there is an important distinction to be made. For now the product $\beta^2F$ no longer has a single limit value at small $\beta$, rather, as we show explicitly in Appendix~\ref{Fderiv}, there is a limit set:
\begin{align}
&\lim_{\beta\rightarrow 0} \varepsilon\beta^2F=
-\sum_{n\in \mathbb{Z}}c_n\, (\beta\varepsilon)^{-\frac{2\pi i n}{\log p}}\,,
\\
\text{with }&c_n=
\frac{(p-1)^2}{\log p}
\frac{\Gamma(1+\frac{2\pi i n}{\log p})\,\zeta(2+\frac{2\pi i n}{\log p})}{(p+1)^{1+\frac{2\pi i n}{\log p}}}\,.
\end{align}
The limit set can alternatively be re-expressed as a Fourier series in $\log(\beta\varepsilon)$: 
\begin{align}
&\hspace{30mm}\lim_{\beta\rightarrow 0} \varepsilon\beta^2F=-c_0+
\\
&
\sum_{n\in\mathbb{N}}\hspace{-0.5mm}\bigg\{\hspace{-0.5mm}A_n\cos\hspace{-0.5mm}\Big(\frac{2\pi }{\log p}\log(\beta\varepsilon)n\Big)
\hspace{-0.5mm}+\hspace{-0.5mm}
B_n\sin\hspace{-0.5mm}\Big(\frac{2\pi }{\log p}\log(\beta\varepsilon)n\Big)\bigg\},
\nonumber \\
&\text{where }A_n=-c_n-c_{-n}
\text{ and }
B_n=i(c_n-c_{-n})\,.
\end{align}
For small values of $p$, the oscillations are highly suppressed, e.g. for $p=2$:
\begin{align}
\frac{A_1}{c_0}\big|_{p=2}\approx 1.418\cdot 10^{-6},
\hspace{3mm}
\frac{B_1}{c_0}\big|_{p=2}
\approx  -7.027\cdot 10^{-6}.
\end{align}
But for larger values of $p$, the oscillations become increasingly pronounced, and in the large $p$ limit one finds that
\begin{align}
\forall n>0: 
\hspace{5mm}
\lim_{p\rightarrow\infty}\frac{A_n}{c_0}=-2\,,
\hspace{10mm}
\lim_{p\rightarrow\infty}\frac{B_n}{c_0}=0\,.
\end{align}
Log-periodic fluctuations are a known phenomenon in the thermodynamics on tree-graphs, see for example \cite{Derrida:1983ty,andrade2000emergence,costin2012oscillatory,Calcagni:2017via,bhoyar2021emergence}. The consequence of such limiting behavior is the result announced in the introduction: the high temperature entropy exhibits a modulated Hardy-Ramanujan type scaling. To see this more precisely, we first write down the expression for the entropy up to zeroth order in $\beta$:
\begin{align}
&S(\beta) \approx \,
\beta E + \frac{A}{\beta \varepsilon}
+B\log(\beta\varepsilon)
-C+
\\ \nonumber
&
\frac{(p-1)^2}{\log p}
\hspace{-1mm}
\sum_{0\neq  n\in \mathbb{Z}}
\hspace{-1mm}
\Gamma(1+\frac{2\pi i n}{\log p})\frac{\zeta(2+\frac{2\pi i n}{\log p})+\frac{2\beta\varepsilon}{p}\zeta(1+\frac{2\pi i n}{\log p})}{\Big((p+1)\beta\varepsilon\Big)^{1+\frac{2\pi i n}{\log p}}}
,
\end{align}
where $A$, $B$, and $C$ are $p$-dependent constants whose precise forms are given in \eqref{Aeq}, \eqref{Beq}, and \eqref{Ceq}. The asymptotic form of the total degeneracy at energy $E$ can again be computed from the entropy via the standard formula that follows from a saddle-point approximation of the partition function,
\begin{align}
\Omega_\text{asym}(E) =\,& \frac{\exp\big[S(\beta_0)\big]}{\sqrt{2\pi S''(\beta_0)}}\,,
\end{align}
where $\beta_0$ is the function of $E$ that solves the saddle point equation $0=S'(\beta_0)$. Equivalently, at large temperatures, $\beta_0(E)$ is the inverse function to
\begin{align}
\label{Ebeta0}
E(\beta_0) = \varepsilon(p-1)^2(p+1)\sum_{k=0}^\infty
\frac{p^{2k}}{e^{\beta_0\varepsilon(p+1)p^k}-1}\,.
\end{align}
We have not found a way to analytically invert the function \eqref{Ebeta0} and for this reason we cannot provide a closed-form expressions for $\Omega_\text{asym}(E)$. However, if we consider the kind of averaged asymptotics that ensues on dropping the log-periodic fluctuations, then we can write down a closed-form answer simply by reusing the earlier equation \eqref{OmegaE} with the modified values of $A$, $B$, and $C$ to find that the coarse-grained overall degeneracy $\overline{\Omega_\text{asym}}(E)$ is given asymptotically by
\begin{align}
\label{OmegaAsymp}
\overline{\Omega_\text{asym}}(E)
\propto \frac{
\exp\Big[
(p-1)\pi\sqrt{\frac{2E/\varepsilon}{3(p+1)\log p}}
\Big]}
{
E^{
\frac{5}{4}
-\frac{(p-1)^2}{p(p+1)\log p}
}
}\,,
\end{align}
where we have omitted a rather lengthy $p$-dependent normalization constant. Figure~\ref{HRplots} shows plots comparing exact degeneracies to the averaged and unaveraged asymptotic formulas.

\subsection{\label{NeumannEntropy}Entropy of Neumann Trees}

In Section~\ref{Neumann}, we reviewed the normal mode spectrum of the infinite $(p+1)$-regular tree with Neumann boundary conditions. We mentioned in the introduction that this spectrum does not produce Hardy-Ramanujan scaling for the degeneracy. The reason for this fact is essentially that while the degeneracy, as shown in equation \eqref{NeumannSpectrum}, grows as $p^n$ like in the ultrametric spectrum, the energy levels grow only as $p^{n/2}$. As the energy levels are less sparse, ultimately there are more ways to piece together energy quanta to obtain any given total energy $E$. More precisely, the asymptotic total degeneracy at large energy $E$ turns out to scale not as $e^{\#\sqrt{E}}$, but rather as $e^{\# E^{2/3}}$. To justify this statement and uncover the coefficient in the exponent, one can first compute the free energy for the system built out of quantum harmonic oscillators associated to the Neumann-tree normal modes, letting $\varepsilon = \hbar \omega_0$, with the result that
\begin{align}
F =\,&
\frac{p}{\beta}\log(1-e^{-\beta\varepsilon})
-\frac{\varepsilon}{p^{1/2}+1+p^{-1/2}}
\\ \nonumber
&
-2\frac{p^2-1}{\beta\,p\log p}
\sum_{n=-\infty}^\infty
\frac{\Gamma(2+\frac{4\pi i n}{\log p})\,\zeta(3+\frac{4\pi i n}{\log p})}{(\beta \varepsilon)^{2+\frac{4\pi i n}{\log p} }}
\\ \nonumber
&
-\frac{p^2-1}{p\beta}
\sum_{n=0}^\infty p^{-n}
\log\Big(1-e^{-\beta\varepsilon p^{-n/2}}\Big)\,.
\end{align}
Again we witness the presence of log-periodic fluctuations, but the envelope $\overline{F}$ has a leading high-temperature behavior given by 
\begin{align}
\overline{F}
= - \frac{2(p^2-1)\zeta(3)}{p \log p\,\beta^3\varepsilon^2} +...\,.
\end{align}
Hence we see that the averaged total energy has a leading scaling given by
\begin{align}
\label{OverlineE}
\overline{E} =\,& \frac{\partial}{\partial \beta}(\beta \overline{F})
\approx 
\frac{4(p^2-1)\zeta(3)}{p \log p\,\beta^3\varepsilon^2}\,,
\end{align}
while the leading scaling for the entropy envelope is given by
\begin{align}
\label{OverlineS}
\overline{S} =\,& \beta^2\frac{\partial}{\partial \beta}\overline{F}
\approx 
\frac{6(p^2-1)\zeta(3)}{p \log p\,\beta^2\varepsilon^2}
\sim \overline{E}^{2/3}\,.
\end{align}
Comparing equations \eqref{OverlineE} and \eqref{OverlineS}, we see that
\begin{align}
\overline{S}
\approx 
3\Big(\frac{(p^2-1)\zeta(3)}{2p\log p}\Big)^{1/3}\,
(\overline{E}/\varepsilon)^{2/3}\,.
\end{align}
This equation makes it clear that, unlike the ultrametric spectrum, the Neumann-tree spectrum behaves qualitatively different from the non-relativistic string, and in this sense it is not appropriate to think of the tree-graph as merely a one-parameter deformation that recovers the usual string when the coordination number of the tree is set to two. While the string free energy scales with temperature squared as an energy density in one spatial dimension the free energy of the tree---even when setting aside the modulations---effectively behaves as an energy density in two spatial dimensions.

\section{\label{Discussion}Discussion}

The core results we have sought to communicate with the present paper consist in the following:
\\[2mm]
1) the precise determination of the eigenvalue spectrum of the Vladimirov derivative when acting on the $p$-adic units $\mathbb{U}_p$, as given in equation \eqref{ultSpectrum}, and also realized as the spectrum of the $p$-adic Neumann-to-Dirichlet operator computed in Appendix~\ref{NeumannToDirichlet}, and
\\[2mm]
2) a saddle-point evaluation of the partition function of the quantum theory built from the infinite set of oscillators with energy quanta matching this spectrum, revealing that, as in string theory, Hardy-Ramanujan scaling is realized, by which we mean that, at large total energy $E$, the total degeneracy scales as $e^{\#\sqrt{E}}$ as shown in equation \eqref{OmegaAsymp}. 
\\[2mm]
An important qualification to the second point, however, is that the leading scaling is modulated by a periodic function in $\log E$ \footnote{While the present work focuses on the entropy and free energy of non-relativistic quantum systems and everything is real, in the context of relativistic CFT correlators, such log-periodic fluctuations would be indicative of a complex CFT \cite{Gorbenko:2018ncu,Gorbenko:2018dtm}.}. But compared to the overall envelope,  the fluctuations are of sufficiently small size that the energy remains a growing function of the temperature and the entropy a concave function of the energy, so that the fluctuations do not compromise thermodynamic stability.

While the ultrametric spectrum that we computed directly in a continuum space of $p$-adic numbers displays stringy microstate growth, we witnessed in Section~\ref{NeumannEntropy} how more crude attempts at constructing string theory deformations by replacing string-graphs with tree-graphs fail to realize this scaling. And in Section~\ref{Periodic} we saw how discretized closed strings occupy a special place in graph theory: they furnish the only infinite family of regular, vertex transitive graphs of maximal girth. In physics parlance, this fact entails that in a hypothetical microscopic model of dynamical graphs that produce smooth objects in a large $N$ limit, string graphs constitute a unique family of translationally invariant objects whose oscillation modes are of a particularly simple analytically tractable kind.

On a mathematical note, the microstate scalings consequent to linearly spaced quantum harmonic oscillator energy quanta are determined generally by a theorem known as Meinardu's theorem \cite{meinardus1953asymptotische}. The types of models we have considered raise the broader question of how this theorem may be expanded, in ways similar to the variations of integer partitioning explored in \cite{hwang2001limit,granovsky2008meinardus,granovsky2012meinardus,latapy2001partitions}, to exponentially spaced and degenerate oscillator energies obtainable from $p$-adic spectral theory.

Another question to address is whether a symmetry property akin to modular invariance is present for the ultrametric partition functions. 
For the non-relativistic string spectrum reviewed in Section~\ref{Non-relativistic}, the modular invariance of the partition function $Z(\beta\varepsilon)=
\sqrt{\frac{\beta\varepsilon}{2\pi}}\,Z(\frac{4\pi^2}{\beta\varepsilon})$ is equivalent to the well-known property of the Dedekind eta function that $\eta(-\frac{1}{\tau})
=\sqrt{-i\tau}\eta(\tau)$. In the high-temperature expansion \eqref{Ffull} for the free energy, modular invariance manifests itself in the property that the power series in $\beta$ terminates but is supplemented by an infinite tower of non-perturbative corrections in $\beta$. In the case of the ultrametric free energy studied in Section~\ref{UltrametricHardRamanujan}, the high temperature expansion \eqref{FexpandFull} of the free energy does not terminate, based on which a notion of modular invariance is not be expected. This fact could be construed as a shortcoming of the models, possibly modified in a refinement thereof, or could simply be taken as a sign that the torus manifold is not applicable to the geometries relevant to the ultrametric models. Besides modular invariance, the integer partition function is host to several important mathematical properties---recurrence relations, modular arithmetic patterns, relations to theta functions, the Rademacher expansion---many with physical applications, and it remains to be determined how many extend to the ultrametric setting.

As mentioned in the introduction, within the community of physicists who study the $S$-matrix bootstrap, the past few years have seen an increase of interest in putative $q$-deformed string amplitudes with an exponentially spaced tower of energy levels, although research in this direction has mostly focused on energy levels with exponentially decreasing spacing, rather than increasing as in the present survey. It may be worth remarking than in such $q$-deformed amplitudes the momentum-dependence is required to enter via the logarithm of Mandelstam invariants. In what is perhaps an unrelated comment, we note here that while the eigenfunctions of the derivative operator on the real circle are an integer-labeled family of \emph{additive characters}, we witnessed in Section~\ref{UltrametricSpectrum} that the eigenfunctions of the Vladimirov derivative on the $p$-adic circle $\mathbb{U}_p$ are the integer-labeled family of \emph{multiplicative characters}.

To further investigate the extent to which the ultrametric spectrum computed in Section~\ref{UltrametricSpectrum} is associated to a genuine and complete physical theory, it will be necessary to develop a more refined understanding of the states associated to these energies, and if possible to work out a set of raising and lowering operators and determine if an underlying notion of Virasoro algebra may be defined for such operators. Unless this step can be performed, it is hard to imagine that $p$-adic string theory can be elevated to a Lorentzian theory with a non-trivial spectrum containing more than a single tachyonic scalar. In the context of $p$-adic AdS/CFT \footnote{The founding references are \cite{Gubser:2016guj} and \cite{Heydeman:2016ldy}, see also \cite{Hung:2019zsk} for its tensor-network implementation and  \cite{Okunishi:2023syy,Okunishi:2024amt,Qu:2024hqq,Huang:2024ilq,Mondal:2025zxx} for more recent work on the subject}, it has been argued \cite{Gubser:2018cha} that the notion of spin is associated to correlator dependencies not just on norms of differences of $p$-adic valued positions but also on some of the digits in the $p$-nary expansion \eqref{xExpansion}. Under this definition, the ultrametric spectrum meets the basic requirement for being associated to states with an infinite tower of spins. But a more refined understanding of the representation theoretic properties of ultrametric states is a necessary prerequisite to developing a full-fledged Lorentzian incarnation of $p$-adic string theory. Such knowledge too is needed to combine vibrational degrees of freedom with translational ditto to obtain the full partition function, which in string theory leads to the discovery of the Hagedorn temperature.

In the context of the purely tachyonic $p$-adic string, the presence of a Hagedorn phase in $p$-adic string theory was reported earlier in \cite{Biswas:2009nx}, which turned on a finite temperature for the effective (real) spacetime string field theory action determined in \cite{Brekke:1987ptq} and discovered a phase with parametrically small ratio of pressure to energy density, and also uncovered evidence for a duality relating the thermal partition function at temperature $T$ to that at $1/T$. We leave it as an open question whether the ultrametric spectrum of the present paper can serve towards expanding the above results by by paving the way for a $p$-adic string field theory analysis encompassing an infinite tower of particles and not the tachyon alone.

Another possible avenue of future inquiry concerns higher-genus geometries, for which an ultrametric formalism phrased in the mathematical language of Mumford curves has been known since the late eighties \cite{Chekhov:1989bg}. Higher-genus amplitudes remain elusive objects in string theory \footnote{An analytic closed-form expression for the one-loop cosmological constant of the bosonic string was written down only last year \cite{Baccianti:2025gll}}, but the discretized setting may offer a means to illuminate their general properties as explicit computations become more readily feasible. The eigenspectrum studied in the present paper can be thought of as describing a set radial characters which can subsequently be combined with a notion of angular characters to produce the building blocks of vacuum amplitudes.

A mathematical classification theorem due to Ostrowski provides a precise sense in which the real and $p$-adic numbers together provide the full range of perspectives that can be brought to bear on anything founded on rational numbers. A possible implication is that there exists a notion of completeness according to which the free string spectrum and the ultrametric spectrum furnish a complete class of integrable systems. Within the context of a sparsely coupled family of spin-chains that interpolate between 2-adic and local coupling, \cite{Bentsen:2019rlr} demonstrated that while the level spacings of the local and ultrametric spin chains followed a Poisson distribution, the spin chain at the midpoint farthest from any notion of locality exhibited Wigner-Dyson level spacings. If this lesson applies more broadly, one could envision the real spectrum and the ultrametric spectra for all $p$ as marking a set of integrable outer endpoints for the convex hull of a space that carries chaos in its interior.

\subsection*{Acknowledgements}

AH and CBJ are grateful to SIMIS for its hospitality during a crucial phase of this work. The work of AH is supported by Simons collaboration grant No. 708790. The work of CBJ is supported by the Korea Institute for Advanced Study (KIAS) Grant No. PG095901.

\appendix

\begin{widetext}
\section{\label{Fderiv}Deriving the full high temperature expansion of the free energy}
In this appendix we explicitly work out the high temperature expansion for the free energy of the ultrametric spectrum that we studied in Section~\ref{UltrametricHardRamanujan}. As a first step we split the free energy $F$ into two pieces,
\begin{align}
F =
\varepsilon F_1(\beta\varepsilon)
+\frac{(p-1)^2}{\beta^2\varepsilon}f(\beta\varepsilon)
\end{align}
where we have defined two functions of $b=\beta \varepsilon$ given by
\begin{align}
F_1(b) \equiv\,&
\frac{p-2}{b}\log\bigg(1-\exp\Big[-b\frac{p-1}{p}\Big]\bigg)\,,
\\ \nonumber
f(b) \equiv \,&
b\sum_{k=1}^\infty
p^{k-1}
\log\bigg(1-\exp\Big[-b \Big(\frac{1+p}{p}p^k-\frac{2}{p}\Big)\Big]\bigg)\,.
\end{align}
Let us first consider $F_1$. The small $b$ expansion here follows from the following identity, where $B_\ell$ are the Bernoulli numbers,
\begin{align}
\label{logId}
\log(1-e^{-x})
=\log x +\sum_{\ell=1}^\infty \frac{B_\ell}{\ell^2\Gamma(\ell)}x^\ell\,.
\end{align}
We therefore have that
\begin{align}
F_1(b) =\,& 
\frac{p-2}{b}\log\Big(b\frac{p-1}{p}\Big)
+\frac{p-2}{b}\sum_{\ell=1}^\infty \frac{B_\ell}{\ell^2\Gamma(\ell)} b^\ell \Big(\frac{p-1}{p}\Big)^\ell
\,.
\end{align}
Next we turn to $f$. This function is given in terms of a semi-infinite sum, but we find it convenient to recast it as a bi-infinite sum minus a correction:
\begin{align}
f(b)=f_1-f_2\,,
\end{align}
with the two terms given as
\begin{align}
f_1\equiv\,b\sum_{k=-\infty}^\infty
p^k
\log\bigg(1-\exp\Big[-b \Big((1+p)p^k-\frac{2}{p}\Big)\Big]\bigg)\,,
\\
\label{f2def}
f_2\equiv\,b\sum_{k=-\infty}^{-1}
p^k
\log\bigg(1-\exp\Big[-b \Big((1+p)p^k-\frac{2}{p}\Big)\Big]\bigg)\,.
\end{align}
The advantage to having a sum over all the integers in $f_1$ is that it permits us to invoke Poission resummation to infer that
\begin{align}
f_1=\,& b\sum_{n=-\infty}^\infty
\int_\mathbb{R}dx\,e^{2\pi i n x}
p^x
\log\bigg(1-\exp\Big[-b \Big((1+p)p^x-\frac{2}{p}\Big)\Big]\bigg)\,.
\end{align}
Expanding the logarithm according to the standard identity
$\log(1-\epsilon)=-\sum_{n=1}^\infty \epsilon^n/n $ and carrying out the $x$ integral, one finds that
\begin{align}
\label{f1sum}
f_1
=\,& -\frac{b}{\log p}\sum_{n=-\infty}^\infty
\sum_{m=1}^\infty \frac{e^{2bm/p}}{m}
\frac{\Gamma\Big(1+\frac{2\pi i n }{\log p}\Big)}{\Big(bm(p+1)\Big)^{1+\frac{2\pi i n}{\log p}}}= -\frac{b}{\log p}\sum_{n=-\infty}^\infty
\text{Li}_{2+\frac{2\pi i n}{\log p}}(e^{2b/p})
\frac{\Gamma\Big(1+\frac{2\pi i n }{\log p}\Big)}{\Big(b(p+1)\Big)^{1+\frac{2\pi i n}{\log p}}}\,,
\end{align}
where we evaluated the sum over $m$ in terms of the polylogarithm function. The advantage to this re-writing is that the small $b$ expansion of $f_1$ then follows from the known expansion of the polylogarithm function. In the case when $n=0$, we need the expansion of the dilogarithm,
\begin{align}
\label{Li0expand}
\text{Li}_2(e^x)=\frac{\pi^2}{6}+x-i\pi x-x\log x+\sum_{m=2}^\infty \frac{B_{m-1}}{(m-1)\,m!}(-x)^m\,,
\end{align}
while for other values of $n$, the polylogarithm has the expansion
\begin{align}
\label{Li2expansion}
\text{Li}_{2+\frac{2\pi i n}{\log p}}(e^x)=\,&
-x^{1+\frac{2\pi i n}{\log p}}e^{-\frac{2\pi^2 n}{\log p}}\Gamma(-1-i\frac{2\pi n}{\log p})
+\sum_{m=0}^\infty \frac{\zeta(2+\frac{2\pi i n}{\log p}-m)}{m!}x^m
\text{ for }n\neq 0\,.
\end{align}
Next let us work out the small $b$ expansion of $f_2$ defined in \eqref{f2def}. For the term with $k=-1$ the argument of the logarithm is positive while we pick up an imaginary piece for other values of $k$:
\begin{align}
\label{f2eval}
f_2=\,&\frac{b}{p}
\log\bigg(1-\exp\Big[-b \Big(\frac{p-1}{p}\Big)\Big]\bigg)
+b\sum_{k=2}^\infty
p^{-k}
\Bigg(i\pi+\log\Big(\exp\Big[b \Big(\frac{2}{p}-(1+p)p^{-k}\Big)\Big]-1\bigg)\Bigg)
\\
=\,& \frac{b}{p}\log b+\frac{b}{p}\log\Big(\frac{p-1}{p}\Big)
+\frac{b}{p}\sum_{n=1}^\infty \frac{B_n}{n^2\Gamma(n)}
b^n\Big(\frac{p-1}{p}\Big)^n
+b\frac{i\pi}{p(p-1)}
+f_{2,2}\,,
\end{align}
where we again made use of \eqref{logId} and the last term is defined by
\begin{align}
f_{2,2}=\,&
b\sum_{k=2}^\infty
p^{-k}
\log\Big(\exp\Big[b \frac{2-(p+1)p^{1-k}}{p}\Big]-1\Big)
\\
=\,&
\frac{b}{p}\sum_{k=1}^\infty
p^{-k}
\Bigg(\log\Big(b \frac{2-(p+1)p^{-k}}{p}\Big)+b \frac{2-(p+1)p^{-k}}{2p}+\sum_{n=2}^\infty\frac{B_n}{n^2\Gamma(n)}\Big(b \frac{2-(p+1)p^{-k}}{p}\Big)^n\Bigg)\,.
\end{align}
Note that in evaluating the negative-argument logarithm in \eqref{f2eval}, we made a choice of branch cut. We chose the cut necessary for the difference of $f_1$ and $f_2$ to produce the real function $f$. For on carrying out the sum over $n$ in \eqref{f1sum}, $f_1$ picks up an imaginary piece owing to the first term on the right-hand side of \eqref{Li2expansion} and this piece, together with an imaginary piece owing to the third term on the right-hand side of \eqref{Li0expand}, precisely cancel the imaginary piece in $f_2$. To see this, we first note that
\begin{align}
e^{-\frac{2\pi^2n}{\log p}}\frac{\Gamma(1+\frac{2\pi i n}{\log p})\,\Gamma(-1-\frac{2\pi i n}{\log p})}{\log p}
=\pi \frac{1-\coth(\frac{2n\pi^2}{\log p})}{2n\pi - i\log p}\,,
\end{align}
and then we observe that 
\begin{align}
b\sum_{n\neq 0}
\frac{\pi}{2n\pi-i\log p}
\Big(\frac{2}{p(p+1)}\Big)^{1+\frac{2\pi i n}{\log p}}= b\frac{\pi i}{p(p-1)}-b\frac{2\pi i}{p(p+1)\log p}\,,
\end{align}
the which two terms exactly cancel the imaginary part of $f_1$.

Having now determined the high temperature expansions for $F_1$, $f_1$, and $f_2$, we find the expansion for the free energy by adding the pieces together according to the identity.
\begin{align}
F=\,&\varepsilon  F_1
+\frac{(p-1)^2}{\beta^2\varepsilon}
(f_1-f_2)\,.
\end{align}
When the dust settles, one finds that
\begin{align}
\label{FexpandFull}
\frac{F}{\varepsilon} =\,& - \frac{1}{b^2}\bigg(A+\sum_{n\in \mathbb{Z}\backslash\{0\}}c_n\,b^{-\frac{2\pi i n}{\log p}} \bigg) -\frac{1}{b}\bigg(
B\, \log b +\sum_{n\in \mathbb{Z}\backslash\{0\}}\widetilde{c}_n\,b^{-\frac{2\pi i n}{\log p}} -C\Bigg) \\ \nonumber
&-\sum_{\ell=0}^\infty
b^\ell\sum_{n\in \mathbb{Z}}
d_{\ell,n}\,b^{-\frac{2\pi i n}{\log p}}
-\sum_{\ell=2}^\infty K_\ell\,b^{\ell-1}
\,, 
\end{align}
where $A$, $B$, $C$,  $c_n$,  $\widetilde{c}_n$, $d_{\ell,n}$, and $K_\ell$ are $p$-independent constants given explicitly by

\begin{align}
\label{Aeq}
A = \frac{\pi^2(p-1)^2}{6(p+1)\log p}\,,
\end{align}

\begin{align}
\label{Beq}
B=1-
\frac{2(p-1)^2}{p(p+1)\log p}\,,
\end{align}

\begin{align}
\label{Ceq}
C =\,&
-\frac{1}{p}\log(\frac{p-1}{p})
-\frac{2(p-1)^2}{p(p+1)}\Big(1+\frac{1-\log 2}{\log p}\Big)
-\frac{(p-1)^2}{p}\sum_{k=1}^\infty p^{-k}\log\Big(\frac{2-(p+1)p^{-k}}{p}\bigg)
\\& \nonumber
- (p-1)^2
\sum_{n\neq 0}
\frac{\pi\coth(\frac{2\pi^2n}{\log p})}{2n\pi-i\log p}
\Big(\frac{2}{p(p+1)}\Big)^{1+\frac{2\pi i n}{\log p}}\,,
\end{align}

\begin{align}
\label{cneq}
c_n=\frac{(p-1)^2}{\log p}
\frac{\Gamma(1+\frac{2\pi i n }{\log p})\zeta(2+\frac{2\pi i n}{\log p})}{(p+1)^{1+\frac{2\pi i n}{\log p}}}\,,
\hspace{10mm}
\widetilde{c}_n=\frac{2(p-1)^2}{p\log p}
\frac{\Gamma(1+\frac{2\pi i n }{\log p})\zeta(1+\frac{2\pi i n}{\log p})}{(p+1)^{1+\frac{2\pi i n}{\log p}}}\,,
\end{align}

\begin{align}
d_{\ell,n}=\frac{4(p-1)^2}{p^2\log p}
\frac{\Gamma(1+\frac{2\pi i n}{\log p})\,\zeta(\frac{2\pi i n}{\log p}-\ell)}{(p+1)^{1+\frac{2\pi i n}{\log p}}}
\frac{(2/p)^\ell}{(2+\ell)!}\,,
\end{align}

\begin{align}
K_\ell = \frac{1}{p}
\bigg(
(p-1)^2\sum_{k=1}^\infty
p^{-k}\big(\frac{2-(p+1)p^{-k}}{p}\big)^\ell
+\big(\frac{p-1}{p}\big)^\ell
\bigg)
\frac{B_\ell}{\ell^2\Gamma(\ell)}\,.
\end{align}
As a practical remark, we note that an obstacle to high precision numerical checks of the expansion
\eqref{FexpandFull} to high order in $b$ is that the determination of the constant $C$ requires the evaluation of the following sum
\begin{align}
\mathcal{S}\equiv\,&
\sum_{n\neq 0}
\frac{\pi\coth(\frac{2\pi^2n}{\log p})}{2n\pi-i\log p}
\Big(\frac{2}{p(p+1)}\Big)^{1+\frac{2\pi i n}{\log p}}
\\ \label{slowSum}
=\,&
\frac{4\pi}{p(p+1)}
\sum_{n=1}^\infty
\coth(\frac{2\pi^2 n}{\log p})
\frac{
2\pi n\cos\big(\frac{2\pi n}{\log p}\log(\frac{2}{p(p+1)})\big)
-\log p \sin\big(\frac{2\pi n}{\log p}\log(\frac{2}{p(p+1)})\big)
}{4\pi^2n^2+(\log p)^2}\,,
\end{align}
which converges slowly. A way to overcome this obstacle is to rewrite the sum by adding and subtracting back a carefully chosen piece:
\begin{align}
\label{AddAndSubtract}
\frac{\coth(\frac{2\pi^2 n}{\log p})}{4\pi^2n^2+(\log p)^2}
=
\bigg(\frac{\coth(\frac{2\pi^2 n}{\log p})}{4\pi^2n^2+(\log p)^2}-\frac{1}{4\pi^2n^2}
+\frac{(\log p)^2}{16\pi^4n^4}
-\frac{(\log p)^4}{64\pi^6n^6}\bigg)+\bigg(\frac{1}{4\pi^2n^2}
-\frac{(\log p)^2}{16\pi^4n^4}
+\frac{(\log p)^4}{64\pi^6n^6}\bigg)\,.
\end{align}
The sum over $n$ in \eqref{slowSum} now converges much faster when carried out over the first piece on the right-hand side of \eqref{AddAndSubtract}, while the sum over the second piece can be evaluated analytically to produce a finite sum over polylogarithms. Including higher order terms in $1/n$ when adding and subtracting in \eqref{AddAndSubtract} further speeds up numerical evaluations.

\end{widetext}

\section{\label{NeumannToDirichlet}The Neumann-to-Dirichlet operator and its eigenvalues}

In this appendix we show that the ultrametric spectrum computed in Section~\ref{UltrametricSpectrum} admits of an alternative interpretation: the eigenfunctions and degeneracies and---up to an overall shift---the eigenvalues as well also provide the eigenspectrum of the Neumann-to-Dirichlet operator $D'$ on $\mathbb{U}_p$. We first define this operator before proceeding to compute its eigenspectrum.

It is a classical result in harmonic analysis that the positive semi-definite square root of the Laplacian on a unit circle, can be interpreted as a Neumann-to-Dirichlet boundary operator: i.e. given a suitable function $f(x)$ on the circle, there is a unique harmonic extension $F(x)$ to the disk. The map from $f(x)$ to the normal derivative of $F(x)$ coincides with the positive semi-definite square root of the Laplacian.

In the following, we define a Neumann-to-Dirichlet boundary operator $D'$ on a leaf of the Bruhat-Tits tree $T_p$, which is a rooted tree $R_p$ with root degree $p-1$, and degree $p+1$ for any other vertex. One has a bijection from the asymptotic boundary of $R_p$: i.e. non-backtracking semi-infinite rays on $R_p$ from the root, to elements of $\mathbb{U}_p$. So $D'$ is to be compared with $D$, the Vladimirov derivative given in equation \eqref{eqDeriv}. 

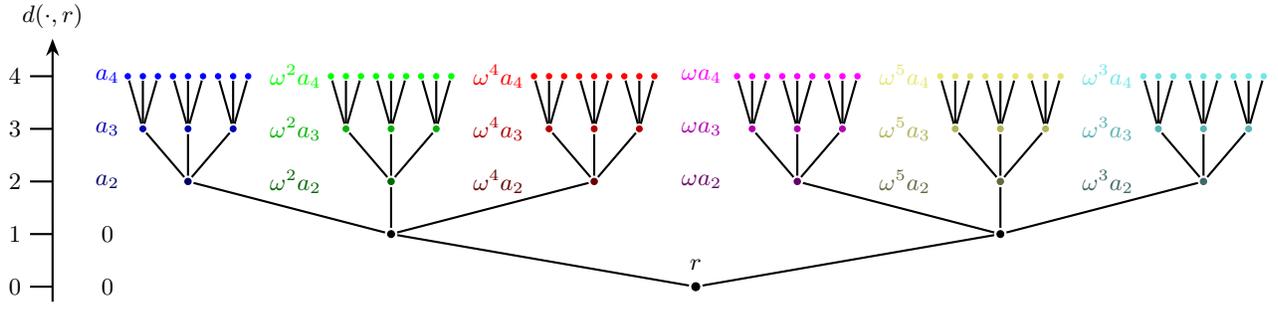
\begin{figure*}
\centering
$
\begin{matrix}
\text{
\scalebox{1}{\begin{tikzpicture}
\node at (3.5+0.7,1.5)  {$d(\cdot,r)$};
\node at (3+0.7,-2.1)  {0};
\node at (3+0.7,-1.4)  {1};
\node at (3+0.7,-0.7)  {2};
\node at (3+0.7,0)  {3};
\node at (3+0.7,0.7)  {4};
\draw [thick,-Stealth](3.5+0.7,-2.3) -- (3.5+0.7,1.2);
\draw [thick](3.2+0.7,-2.1) -- (3.5+0.7,-2.1);
\draw [thick](3.2+0.7,-1.4) -- (3.5+0.7,-1.4);
\draw [thick](3.2+0.7,-0.7) -- (3.5+0.7,-0.7);
\draw [thick](3.2+0.7,0) -- (3.5+0.7,0);
\draw [thick](3.2+0.7,0.7) -- (3.5+0.7,0.7);
\draw [thick](5.4,0) -- (5.2,0.7);
\draw [thick](5.4,0) -- (5.4,0.7);
\draw [thick](5.4,0) -- (5.6,0.7);
\draw [thick](6,0) -- (5.8,0.7);
\draw [thick](6,0) -- (6,0.7);
\draw [thick](6,0) -- (6.2,0.7);
\draw [thick](6.6,0) -- (6.4,0.7);
\draw [thick](6.6,0) -- (6.6,0.7);
\draw [thick](6.6,0) -- (6.8,0.7);
\draw [thick](6,-0.7) -- (5.4,0);
\draw [thick](6,-0.7) -- (6,0);
\draw [thick](6,-0.7) -- (6.6,0);
\draw [thick](5.4+2.7,0) -- (5.2+2.7,0.7);
\draw [thick](5.4+2.7,0) -- (5.4+2.7,0.7);
\draw [thick](5.4+2.7,0) -- (5.6+2.7,0.7);
\draw [thick](6+2.7,0) -- (5.8+2.7,0.7);
\draw [thick](6+2.7,0) -- (6+2.7,0.7);
\draw [thick](6+2.7,0) -- (6.2+2.7,0.7);
\draw [thick](6.6+2.7,0) -- (6.4+2.7,0.7);
\draw [thick](6.6+2.7,0) -- (6.6+2.7,0.7);
\draw [thick](6.6+2.7,0) -- (6.8+2.7,0.7);
\draw [thick](6+2.7,-0.7) -- (5.4+2.7,0);
\draw [thick](6+2.7,-0.7) -- (6+2.7,0);
\draw [thick](6+2.7,-0.7) -- (6.6+2.7,0);
\draw [thick](5.4+5.4,0) -- (5.2+5.4,0.7);
\draw [thick](5.4+5.4,0) -- (5.4+5.4,0.7);
\draw [thick](5.4+5.4,0) -- (5.6+5.4,0.7);
\draw [thick](6+5.4,0) -- (5.8+5.4,0.7);
\draw [thick](6+5.4,0) -- (6+5.4,0.7);
\draw [thick](6+5.4,0) -- (6.2+5.4,0.7);
\draw [thick](6.6+5.4,0) -- (6.4+5.4,0.7);
\draw [thick](6.6+5.4,0) -- (6.6+5.4,0.7);
\draw [thick](6.6+5.4,0) -- (6.8+5.4,0.7);
\draw [thick](6+5.4,-0.7) -- (5.4+5.4,0);
\draw [thick](6+5.4,-0.7) -- (6+5.4,0);
\draw [thick](6+5.4,-0.7) -- (6.6+5.4,0);
\draw [thick](5.4+8.1,0) -- (5.2+8.1,0.7);
\draw [thick](5.4+8.1,0) -- (5.4+8.1,0.7);
\draw [thick](5.4+8.1,0) -- (5.6+8.1,0.7);
\draw [thick](6+8.1,0) -- (5.8+8.1,0.7);
\draw [thick](6+8.1,0) -- (6+8.1,0.7);
\draw [thick](6+8.1,0) -- (6.2+8.1,0.7);
\draw [thick](6.6+8.1,0) -- (6.4+8.1,0.7);
\draw [thick](6.6+8.1,0) -- (6.6+8.1,0.7);
\draw [thick](6.6+8.1,0) -- (6.8+8.1,0.7);
\draw [thick](6+8.1,-0.7) -- (5.4+8.1,0);
\draw [thick](6+8.1,-0.7) -- (6+8.1,0);
\draw [thick](6+8.1,-0.7) -- (6.6+8.1,0);
\draw [thick](5.4+10.8,0) -- (5.2+10.8,0.7);
\draw [thick](5.4+10.8,0) -- (5.4+10.8,0.7);
\draw [thick](5.4+10.8,0) -- (5.6+10.8,0.7);
\draw [thick](6+10.8,0) -- (5.8+10.8,0.7);
\draw [thick](6+10.8,0) -- (6+10.8,0.7);
\draw [thick](6+10.8,0) -- (6.2+10.8,0.7);
\draw [thick](6.6+10.8,0) -- (6.4+10.8,0.7);
\draw [thick](6.6+10.8,0) -- (6.6+10.8,0.7);
\draw [thick](6.6+10.8,0) -- (6.8+10.8,0.7);
\draw [thick](6+10.8,-0.7) -- (5.4+10.8,0);
\draw [thick](6+10.8,-0.7) -- (6+10.8,0);
\draw [thick](6+10.8,-0.7) -- (6.6+10.8,0);
\draw [thick](5.4+13.5,0) -- (5.2+13.5,0.7);
\draw [thick](5.4+13.5,0) -- (5.4+13.5,0.7);
\draw [thick](5.4+13.5,0) -- (5.6+13.5,0.7);
\draw [thick](6+13.5,0) -- (5.8+13.5,0.7);
\draw [thick](6+13.5,0) -- (6+13.5,0.7);
\draw [thick](6+13.5,0) -- (6.2+13.5,0.7);
\draw [thick](6.6+13.5,0) -- (6.4+13.5,0.7);
\draw [thick](6.6+13.5,0) -- (6.6+13.5,0.7);
\draw [thick](6.6+13.5,0) -- (6.8+13.5,0.7);
\draw [thick](6+13.5,-0.7) -- (5.4+13.5,0);
\draw [thick](6+13.5,-0.7) -- (6+13.5,0);
\draw [thick](6+13.5,-0.7) -- (6.6+13.5,0);
\draw [thick](8.7,-1.4) -- (6,-0.7);
\draw [thick](8.7,-1.4) -- (8.7,-0.7);
\draw [thick](8.7,-1.4) -- (11.4,-0.7);
\draw [thick](8.7+8.1,-1.4) -- (6+8.1,-0.7);
\draw [thick](8.7+8.1,-1.4) -- (8.7+8.1,-0.7);
\draw [thick](8.7+8.1,-1.4) -- (11.4+8.1,-0.7);
\draw [thick](12.75,-2.1) -- (8.7,-1.4);
\draw [thick](12.75,-2.1) -- (16.8,-1.4);
\filldraw[white] (12.75,-2.1) circle (2.4pt);
\filldraw[black] (12.75,-2.1) circle (1.5pt);
\node at (12.75,-1.8)  {$r$};
\node at (4.9+0.03,0.7)  {$\textcolor{blue3}{a_4}$};
\node at (4.9+0.03,0)  {$\textcolor{blue2}{a_3}$};
\node at (4.9+0.03,-0.7)  {$\textcolor{blue1}{a_2}$};
\node at (4.9+0.03,-1.4)  {$0$};
\node at (4.9+0.03,-2.1)  {$0$};
\node at (4.8+2.6+0.03,0.7)  {$\textcolor{green3}{\omega^2a_4}$};
\node at (4.8+2.6+0.03,0)  {$\textcolor{green2}{\omega^2a_3}$};
\node at (4.8+2.6+0.03,-0.7)  {$\textcolor{green1}{\omega^2a_2}$};
\node at (4.8+5.3+0.03,0.7)  {$\textcolor{red3}{\omega^4a_4}$};
\node at (4.8+5.3+0.03,0)  {$\textcolor{red2}{\omega^4a_3}$};
\node at (4.8+5.3+0.03,-0.7)  {$\textcolor{red1}{\omega^4a_2}$};
\node at (4.8+8+0.03,0.7)  {$\textcolor{purple3}{\omega a_4}$};
\node at (4.8+8+0.03,0)  {$\textcolor{purple2}{\omega a_3}$};
\node at (4.8+8+0.03,-0.7)  {$\textcolor{purple1}{\omega a_2}$};
\node at (4.8+10.7+0.03,0.7)  {$\textcolor{yellow3}{\omega^5a_4}$};
\node at (4.8+10.7+0.03,0)  {$\textcolor{yellow2}{\omega^5a_3}$};
\node at (4.8+10.7+0.03,-0.7)  {$\textcolor{yellow1}{\omega^5a_2}$};
\node at (4.8+13.4+0.03,0.7)  {$\textcolor{cyan3}{\omega^3a_4}$};
\node at (4.8+13.4+0.03,0)  {$\textcolor{cyan2}{\omega^3a_3}$};
\node at (4.8+13.4+0.03,-0.7)  {$\textcolor{cyan1}{\omega^3a_2}$};
\filldraw[white] (5.2,0.7) circle (1.5pt);
\filldraw[blue3] (5.2,0.7) circle (1pt);
\filldraw[white] (5.4,0.7) circle (1.5pt);
\filldraw[blue3] (5.4,0.7) circle (1pt);
\filldraw[white] (5.6,0.7) circle (1.5pt);
\filldraw[blue3] (5.6,0.7) circle (1pt);
\filldraw[white] (5.8,0.7) circle (1.5pt);
\filldraw[blue3] (5.8,0.7) circle (1pt);
\filldraw[white] (6.0,0.7) circle (1.5pt);
\filldraw[blue3] (6.0,0.7) circle (1pt);
\filldraw[white] (6.2,0.7) circle (1.5pt);
\filldraw[blue3] (6.2,0.7) circle (1pt);
\filldraw[white] (6.4,0.7) circle (1.5pt);
\filldraw[blue3] (6.4,0.7) circle (1pt);
\filldraw[white] (6.6,0.7) circle (1.5pt);
\filldraw[blue3] (6.6,0.7) circle (1pt);
\filldraw[white] (6.8,0.7) circle (1.5pt);
\filldraw[blue3] (6.8,0.7) circle (1pt);
\filldraw[white] (5.2+2.7,0.7) circle (1.5pt);
\filldraw[green3] (5.2+2.7,0.7) circle (1pt);
\filldraw[white] (5.4+2.7,0.7) circle (1.5pt);
\filldraw[green3] (5.4+2.7,0.7) circle (1pt);
\filldraw[white] (5.6+2.7,0.7) circle (1.5pt);
\filldraw[green3] (5.6+2.7,0.7) circle (1pt);
\filldraw[white] (5.8+2.7,0.7) circle (1.5pt);
\filldraw[green3] (5.8+2.7,0.7) circle (1pt);
\filldraw[white] (6.0+2.7,0.7) circle (1.5pt);
\filldraw[green3] (6.0+2.7,0.7) circle (1pt);
\filldraw[white] (6.2+2.7,0.7) circle (1.5pt);
\filldraw[green3] (6.2+2.7,0.7) circle (1pt);
\filldraw[white] (6.4+2.7,0.7) circle (1.5pt);
\filldraw[green3] (6.4+2.7,0.7) circle (1pt);
\filldraw[white] (6.6+2.7,0.7) circle (1.5pt);
\filldraw[green3] (6.6+2.7,0.7) circle (1pt);
\filldraw[white] (6.8+2.7,0.7) circle (1.5pt);
\filldraw[green3] (6.8+2.7,0.7) circle (1pt);
\filldraw[white] (5.2+5.4,0.7) circle (1.5pt);
\filldraw[red3] (5.2+5.4,0.7) circle (1pt);
\filldraw[white] (5.4+5.4,0.7) circle (1.5pt);
\filldraw[red3] (5.4+5.4,0.7) circle (1pt);
\filldraw[white] (5.6+5.4,0.7) circle (1.5pt);
\filldraw[red3] (5.6+5.4,0.7) circle (1pt);
\filldraw[white] (5.8+5.4,0.7) circle (1.5pt);
\filldraw[red3] (5.8+5.4,0.7) circle (1pt);
\filldraw[white] (6.0+5.4,0.7) circle (1.5pt);
\filldraw[red3] (6.0+5.4,0.7) circle (1pt);
\filldraw[white] (6.2+5.4,0.7) circle (1.5pt);
\filldraw[red3] (6.2+5.4,0.7) circle (1pt);
\filldraw[white] (6.4+5.4,0.7) circle (1.5pt);
\filldraw[red3] (6.4+5.4,0.7) circle (1pt);
\filldraw[white] (6.6+5.4,0.7) circle (1.5pt);
\filldraw[red3] (6.6+5.4,0.7) circle (1pt);
\filldraw[white] (6.8+5.4,0.7) circle (1.5pt);
\filldraw[red3] (6.8+5.4,0.7) circle (1pt);
\filldraw[white] (5.2+8.1,0.7) circle (1.5pt);
\filldraw[purple3] (5.2+8.1,0.7) circle (1pt);
\filldraw[white] (5.4+8.1,0.7) circle (1.5pt);
\filldraw[purple3] (5.4+8.1,0.7) circle (1pt);
\filldraw[white] (5.6+8.1,0.7) circle (1.5pt);
\filldraw[purple3] (5.6+8.1,0.7) circle (1pt);
\filldraw[white] (5.8+8.1,0.7) circle (1.5pt);
\filldraw[purple3] (5.8+8.1,0.7) circle (1pt);
\filldraw[white] (6.0+8.1,0.7) circle (1.5pt);
\filldraw[purple3] (6.0+8.1,0.7) circle (1pt);
\filldraw[white] (6.2+8.1,0.7) circle (1.5pt);
\filldraw[purple3] (6.2+8.1,0.7) circle (1pt);
\filldraw[white] (6.4+8.1,0.7) circle (1.5pt);
\filldraw[purple3] (6.4+8.1,0.7) circle (1pt);
\filldraw[white] (6.6+8.1,0.7) circle (1.5pt);
\filldraw[purple3] (6.6+8.1,0.7) circle (1pt);
\filldraw[white] (6.8+8.1,0.7) circle (1.5pt);
\filldraw[purple3] (6.8+8.1,0.7) circle (1pt);
\filldraw[white] (5.2+10.8,0.7) circle (1.5pt);
\filldraw[yellow3] (5.2+10.8,0.7) circle (1pt);
\filldraw[white] (5.4+10.8,0.7) circle (1.5pt);
\filldraw[yellow3] (5.4+10.8,0.7) circle (1pt);
\filldraw[white] (5.6+10.8,0.7) circle (1.5pt);
\filldraw[yellow3] (5.6+10.8,0.7) circle (1pt);
\filldraw[white] (5.8+10.8,0.7) circle (1.5pt);
\filldraw[yellow3] (5.8+10.8,0.7) circle (1pt);
\filldraw[white] (6.0+10.8,0.7) circle (1.5pt);
\filldraw[yellow3] (6.0+10.8,0.7) circle (1pt);
\filldraw[white] (6.2+10.8,0.7) circle (1.5pt);
\filldraw[yellow3] (6.2+10.8,0.7) circle (1pt);
\filldraw[white] (6.4+10.8,0.7) circle (1.5pt);
\filldraw[yellow3] (6.4+10.8,0.7) circle (1pt);
\filldraw[white] (6.6+10.8,0.7) circle (1.5pt);
\filldraw[yellow3] (6.6+10.8,0.7) circle (1pt);
\filldraw[white] (6.8+10.8,0.7) circle (1.5pt);
\filldraw[yellow3] (6.8+10.8,0.7) circle (1pt);
\filldraw[white] (5.2+13.5,0.7) circle (1.5pt);
\filldraw[cyan3] (5.2+13.5,0.7) circle (1pt);
\filldraw[white] (5.4+13.5,0.7) circle (1.5pt);
\filldraw[cyan3] (5.4+13.5,0.7) circle (1pt);
\filldraw[white] (5.6+13.5,0.7) circle (1.5pt);
\filldraw[cyan3] (5.6+13.5,0.7) circle (1pt);
\filldraw[white] (5.8+13.5,0.7) circle (1.5pt);
\filldraw[cyan3] (5.8+13.5,0.7) circle (1pt);
\filldraw[white] (6.0+13.5,0.7) circle (1.5pt);
\filldraw[cyan3] (6.0+13.5,0.7) circle (1pt);
\filldraw[white] (6.2+13.5,0.7) circle (1.5pt);
\filldraw[cyan3] (6.2+13.5,0.7) circle (1pt);
\filldraw[white] (6.4+13.5,0.7) circle (1.5pt);
\filldraw[cyan3] (6.4+13.5,0.7) circle (1pt);
\filldraw[white] (6.6+13.5,0.7) circle (1.5pt);
\filldraw[cyan3] (6.6+13.5,0.7) circle (1pt);
\filldraw[white] (6.8+13.5,0.7) circle (1.5pt);
\filldraw[cyan3] (6.8+13.5,0.7) circle (1pt);
\filldraw[white] (5.4,0) circle (1.8pt);
\filldraw[blue2] (5.4,0) circle (1.1pt);
\filldraw[white] (6,0) circle (1.8pt);
\filldraw[blue2] (6,0) circle (1.1pt);
\filldraw[white] (6.6,0) circle (1.8pt);
\filldraw[blue2] (6.6,0) circle (1.1pt);
\filldraw[white] (8.1,0) circle (1.8pt);
\filldraw[green2] (8.1,0) circle (1.1pt);
\filldraw[white] (8.7,0) circle (1.8pt);
\filldraw[green2] (8.7,0) circle (1.1pt);
\filldraw[white] (9.3,0) circle (1.8pt);
\filldraw[green2] (9.3,0) circle (1.1pt);
\filldraw[white] (10.8,0) circle (1.8pt);
\filldraw[red2] (10.8,0) circle (1.1pt);
\filldraw[white] (11.4,0) circle (1.8pt);
\filldraw[red2] (11.4,0) circle (1.1pt);
\filldraw[white] (12,0) circle (1.8pt);
\filldraw[red2] (12,0) circle (1.1pt);
\filldraw[white] (13.5,0) circle (1.8pt);
\filldraw[purple2] (13.5,0) circle (1.1pt);
\filldraw[white] (14.1,0) circle (1.8pt);
\filldraw[purple2] (14.1,0) circle (1.1pt);
\filldraw[white] (14.7,0) circle (1.8pt);
\filldraw[purple2] (14.7,0) circle (1.1pt);
\filldraw[white] (16.2,0) circle (1.8pt);
\filldraw[yellow2] (16.2,0) circle (1.1pt);
\filldraw[white] (16.8,0) circle (1.8pt);
\filldraw[yellow2] (16.8,0) circle (1.1pt);
\filldraw[white] (17.4,0) circle (1.8pt);
\filldraw[yellow2] (17.4,0) circle (1.1pt);
\filldraw[white] (18.9,0) circle (1.8pt);
\filldraw[cyan2] (18.9,0) circle (1.1pt);
\filldraw[white] (19.5,0) circle (1.8pt);
\filldraw[cyan2] (19.5,0) circle (1.1pt);
\filldraw[white] (20.1,0) circle (1.8pt);
\filldraw[cyan2] (20.1,0) circle (1.1pt);
\filldraw[white] (6,-0.7) circle (2pt);
\filldraw[blue1] (6,-0.7) circle (1.2pt);
\filldraw[white] (8.7,-0.7) circle (2pt);
\filldraw[green1] (8.7,-0.7) circle (1.2pt);
\filldraw[white] (11.4,-0.7) circle (2pt);
\filldraw[red1] (11.4,-0.7) circle (1.2pt);
\filldraw[white] (14.1,-0.7) circle (2pt);
\filldraw[purple1] (14.1,-0.7) circle (1.2pt);
\filldraw[white] (16.8,-0.7) circle (2pt);
\filldraw[yellow1] (16.8,-0.7) circle (1.2pt);
\filldraw[white] (19.5,-0.7) circle (2pt);
\filldraw[cyan1] (19.5,-0.7) circle (1.2pt);
\filldraw[white] (8.7,-1.4) circle (2.2pt);
\filldraw[black] (8.7,-1.4) circle (1.3pt);
\filldraw[white] (16.8,-1.4) circle (2.2pt);
\filldraw[black] (16.8,-1.4) circle (1.3pt);
\end{tikzpicture}}
}\end{matrix}$
\caption{
Harmonic extension $F_l(x)$ of a multiplicative character of conductor $l=2$ for the case of the truncated finite tree $R_p(N)$ with $p=3$ and $N=4$. Whenever $x$ is a vertex whose distance from the root vertex $r$ is less than $l$, we have that $f(x)=0$. In this example, $\omega=e^{\frac{2\pi i }{6}j}$ where $j\in\{1,2,4,5\}$. In our normalization convention $a_4=1$ while $a_2$ and $a_3$ are determined by the equations $(p+1)a_3=pa_4+a_2$ and $(p+1)a_2=pa_3$.
\label{DirichletToNeumann} 
} 
\end{figure*}

To define $D'$, we first truncate $R_p$ by a radius $N\in \mathbb{N}$: i.e. distance from the root. We denote the truncated finite tree by $R_p(N)$. See Figure~\ref{DirichletToNeumann}  for an example. For any $N$, the boundary of $R_p(N)$ is identified as the outmost layer of $R_p(N)$, i.e. the collection of vertexes with distance $N$ from the root. Given any function $f(x)$ on the boundary, there is a unique harmonic extension of it to $R_p$, denoted by $F(x)$. Now, for any vertex $x$ on the boundary, there is a unique neighbor $x'$ of $x$ on $R_p(N)$. As in Zabrodin \cite{Zabrodin:1988ep}, we define the normal derivative of $F(x)$ at $x$ by $p^{N-1}(F(x)-F(x'))$. The linear operator $D'$ on the space of functions on the boundary of $R_p(N)$, is then defined by mapping $f(x)$ to the normal derivative of $F(x)$.

Next, we show that continuous characters of $\mathbb{U}_p$ with conductor $\leq N$ diagonalize $D'$ on $R_p(N)$: i.e. they are eigenvectors of $D'$, and they form a basis of the space of functions on the boundary of $R_p(N)$.

Let $\xi_l$ be such a character with conductor $l\leq N$. The boundary of $R_p(N)$ is identified with the quotient $\mathbb{U}_p/(1+p^N\mathbb{Z}_p)$. One has the obvious quotient map $\mathbb{U}_p/(1+p^N\mathbb{Z}_p)\to \mathbb{U}_p/(1+p^l\mathbb{Z}_p)$. So $\xi_l$ gives rise to a function $f_l(x)$ on the boundary of $R_p(N)$ by the pull-back. Denote by $F_l(x)$ the unique harmonic extension of $f_l(x)$ to $R_p(N)$. Denote the root by $r$. We have the following simple observations:

First, if $l=N$, then for each vertex $x'$ in the 2nd outmost layer of $R_p(N)$, i.e. a vertex with distance $N-1$ from the root, $\sum_{x\sim x', d(x,r)=N} F_l(x)=0$. This is because when $x'$ is the identity element of $\mathbb{Z}_p^*/(1+p^{N-1}\mathbb{Z}_p)$, its neighbors on the boundary of $R_p(N)$ consist of the kernel of the quotient map $\mathbb{U}_p/(1+p^{N}\mathbb{Z}_p)\to \mathbb{U}_p/(1+p^{N-1}\mathbb{Z}_p)$. For any other $x'$, its neighbors on the boundary of $R_p(N)$ is a shift of the kernel of the quotient map by a group element. $\xi_l$ can not restrict to a constant on the kernel, since if that's the case, then it restricts to a constant on the set of neighbors on the boundary of $R_p(N)$ for any $x'$, which would then force the conductor to be less than $l$. Therefore $\xi_l$ restricts to a non-trivial character on the kernel, which implies our observation.   

Second, if $l<N$, then for each vertex $x'$ in the 2nd outmost layer of $R_p(N)$, $F_l(x)$ restricts to a constant. In this case, let $y'$ be a vertex with $d(r,y')=l-1$, we have $\sum_{y\sim y', d(y,r)=l} F_l(y)=0$. This is because for any $y$ with $d(y,r)=l$, $F_l(y)=cF_l(x)$ where $x$ is a vertex with $d(x,r)=N$, and $x$ lies on the same rooted ray as $y$, where $c$ is a nonzero constant independent of $y$. This will become apparent once we perform a computation below for the harmonic extension. For the same reason as above, $F_l$ can not restrict to a constant on neighbors of $y'$ with distance $l$ from the root. So our claim follows.

Next, assume $1\leq l\leq N$. We perform a calculation of the harmonic extension up to the $(l-1)$-th layer of $R_p(N)$. In fact, we only perform the computation for the portion of $R_p(N)$ that is extended back from the identity element of $\mathbb{U}_p/(1+p^{N-1}\mathbb{Z}_p)$, up to the $(l-1)$-th layer of $R_p(N)$. The other cases will follow obviously.

For this, we denote by $a_k$ the value of $F$ along this portion on vertices with distance $k$ from the root. By the definition of harmonic extension, we have
\begin{equation}
\label{rec}
   (p+1)a_{k-1}=a_{k-2}+pa_k \,,
\end{equation}
where $l\leq k\leq N$. By definition, $a_N=1$. By our second observation, we have $a_{l-1}=0$. In fact, $F$ takes zero value on any vertex with distance less than $l$ to the root. These boundary conditions together with equation \eqref{rec} stipulate a recursive problem whose solution can straightforwardly be determined to be
\begin{align}
a_k = \frac{1-p^{l-k-1}}{1-p^{l-N-1}}  \hspace{4mm}\text{ for }\hspace{4mm}l-1 \leq k \leq N\,.
\end{align}

From the foregoing, we see clearly that $f(x)$ is an eigenfunction of $D'$. Denote the eigenvalue by $\lambda_{N,l}$. We have
\begin{equation}\label{eigenvalue}
\lambda_{N,l}=p^{N-1}(1-a_{N-1})%=p^{l-1}a_l
=p^{l-1}\frac{1-p^{-1}}{1-p^{-(N-l+1)}}\,.
\end{equation}
Now, letting $N\to\infty$, the eigenvalue has a limit given by $p^{l-1}(1-p^{-1})$.

Finally, the constant character lies obviously in the kernel of $D'$. So in conclusion, we see that $D'$ and $D$ share the eigenvectors and their multiplicities. For both operators, the eigenvalues depend only on the conductor. Their eigenvalues also share the same leading power dependence on $p$, yet the eigenvalues are different. But the difference amounts only to an overall shift and a choice of normalization.

For this spectrum, the high-temperature entropy again exhibits Hardy-Ramanujan scaling, though subleading terms differ from the thermodynamics of the Vladimirov eigenspectrum given in \eqref{ultSpectrum}. To be more precise, we have just determined the Neumann-to-Dirichlet spectrum to be given---in addition to a zero eigenvalue whose eigenfunction is the trivial character---by
\begin{align}
&E_n = \varepsilon\, p^n \hspace{5mm}\text{for}\hspace{5mm}n\in\mathbb{N}_0\,,
\\ \nonumber
& \rho(0)=p-2\,,
\hspace{5mm}
\rho(n>0)=(p-1)^2p^{n-1}\,,
\end{align}
where $E_n$ is the eigenvalue whose $\rho(n)$ eigenfunctions are given by the multiplicative characters of conductor $n+1$. On associating a quantum harmonic oscillator to each normal mode, the resulting partition function---including the zero-point energy---is given by
\begin{align}
Z =\,& 
\prod_{k=0}^\infty
\bigg(\sum_{\ell=0}^\infty e^{-\beta E_k (\ell+\frac{1}{2})}\bigg)^{\rho(k)}\,.
\end{align}
Regulating the zero-point energy with geometric summation as in \eqref{zeroPointEnergy}, the ground state is now tachyonic:
\begin{align}
\nonumber
\sum_{k=0}^\infty
\frac{\rho(k)\,E_k}{2}
=\,& 
\frac{\varepsilon}{2}
\bigg(
(p-2)+\sum_{n=1}^\infty  (p-1)^2p^{n-1}\,p^n
\bigg)
\\=\,&-\frac{\varepsilon}{p+1}\,.
\end{align}
By a simpler version of the same steps used in Appendix~\ref{Fderiv} and letting $b=\beta\varepsilon$ again, the full high-temperature expansion of the free energy can be worked out with the result that:
\begin{align}
\frac{F}{\varepsilon}=&-\frac{1}{b}\log Z
\\ \nonumber
=&
-\frac{(p-1)^2}{p \log p}\hspace{-1mm}
\sum_{n=-\infty}^\infty
\hspace{-1mm}
\frac{\zeta(2+\frac{2\pi i n}{\log p})\Gamma(1+\frac{2\pi i n}{\log p})}{b^{2+\frac{2\pi i n}{\log p}}}
- \frac{\log (b/p)}{b}
\\
&
-
\sum_{\ell=2}^\infty
\frac{B_\ell}{\ell^2\Gamma(\ell)}
\frac{p^l+p-2}{p^{l+1}-1}
b^{\ell-1}\,.
\end{align}
Incidentally, the zero-point energy has precisely such a value that $F$ contains no term of order $b^0$. Again we observe a leading $F \sim b^{-2}$ scaling modulated by log-periodic fluctuations. By a saddle point approximation of the partition function, one finds that in this case the coarse-grained degeneracy that provides the envelope about which the fluctuations oscillate is given by:
\begin{align}
\overline{\Omega_\text{asymp}}(E)=\,&
\frac{1}{\varepsilon}\frac{\pi}{2p}\Big(\frac{(p-1)^2}{6p\log p}\Big)^{3/4}
\Big(\frac{\varepsilon}{E}\Big)^{\frac{5}{4}} \times
\\ & \nonumber
\exp \Big[
\pi (p-1)\sqrt{\frac{2}{3p\log p}\frac{E}{\varepsilon}}
\Big]\,,
\end{align}
with the leading exponent being identical to that of equation \eqref{OmegaAsymp} but with subleading factors differing.

\bibliography{literature}

\end{document}